\documentstyle[12pt]{article}

\def\@citex[#1]#2{%
\if@filesw \immediate \write \@auxout {\string \citation {#2}}\fi
\@tempcntb\m@ne \let\@h@ld\relax \def\@citea{}%
\@cite{%
  \@for \@citeb:=#2\do {%
    \@ifundefined {b@\@citeb}%
      {\@h@ld\@citea\@tempcntb\m@ne{\bf ?}%
      \@warning {Citation `\@citeb ' on page \thepage \space undefined}}%
      {\@tempcnta\@tempcntb \advance\@tempcnta\@ne%
      \@tempcntb\number\csname b@\@citeb \endcsname \relax%
      \ifnum\@tempcnta=\@tempcntb 
	\ifx\@h@ld\relax%
	  \edef \@h@ld{\@citea\csname b@\@citeb\endcsname}%
	\else%
	  \edef\@h@ld{\ifmmode{-}\else--\fi\csname b@\@citeb\endcsname}%
	\fi%
      \else
	\@h@ld\@citea\csname b@\@citeb \endcsname%
	\let\@h@ld\relax%
      \fi}%
    \def\@citea{,\penalty\@highpenalty\,}%
  }\@h@ld
}{#1}}
\newcount\hour
\newcount\minute
\newtoks\amorpm
\hour=\time\divide\hour by 60
\minute=\time{\multiply\hour by 60 \global\advance\minute by-\hour}
\edef\standardtime{{\ifnum\hour<12 \global\amorpm={am}%
	\else\global\amorpm={pm}\advance\hour by-12 \fi
	\ifnum\hour=0 \hour=12 \fi
	\number\hour:\ifnum\minute<10 0\fi\number\minute\the\amorpm}}
\edef\militarytime{\number\hour:\ifnum\minute<10 0\fi\number\minute}

\def\draftlabel#1{{\@bsphack\if@filesw {\let\thepage\relax
   \xdef\@gtempa{\write\@auxout{\string
      \newlabel{#1}{{\@currentlabel}{\thepage}}}}}\@gtempa
   \if@nobreak \ifvmode\nobreak\fi\fi\fi\@esphack}
	\gdef\@eqnlabel{#1}}
\def\@eqnlabel{}
\def\@vacuum{}
\def\marginnote#1{}
\def\draftmarginnote#1{\marginpar{\raggedright\scriptsize\tt#1}}

\def\draft{\oddsidemargin -.5truein
	\def\@oddhead{\sl \phantom{\today\quad\militarytime} \hfil
	\smash{\Large\sl DRAFT} \hfil \today\quad\militarytime}
	\let\@evenhead\@oddhead
	\let\label=\draftlabel
	\let\marginnote=\draftmarginnote
	\def\ps@empty{\let\@mkboth\@gobbletwo
	\def\@oddfoot{\hfil \smash{\Large\sl DRAFT} \hfil}
	\let\@evenfoot\@oddhead}
	\def\@eqnnum{(\theequation)\rlap{\kern\marginparsep\tt\@eqnlabel}%
	\global\let\@eqnlabel\@vacuum}  }

\def\nblack{            
\def\ZZ{Z \n{10} Z}
\def\NN{N \n{14} N}
\def\CC{C \n{11} C}
\def\RR{R \n{11} R}
\def\QQ{Q \n{12} Q}
\def\PP{P \n{11} P}
}

\def\eqalign#1{\null\,\vcenter{\openup\jot\m@th
  \ialign{\strut\hfil$\displaystyle{##}$&$\displaystyle{{}##}$\hfil
      \crcr#1\crcr}}\,}
\def\eqalignno#1{\displ@y \tabskip\centering
  \halign to\displaywidth{\hfil$\@lign\displaystyle{##}$\tabskip\z@skip
    &$\@lign\displaystyle{{}##}$\hfil\tabskip\centering
    &\llap{$\@lign##$}\tabskip\z@skip\crcr
    #1\crcr}}

\catcode`\@=12

\def\beq{\begin{equation}}
\def\eeq{\end{equation}}
\def\beqar{\begin{eqnarray}}
\def\eeqar{\end{eqnarray}}

\def\nfrac#1#2{{\displaystyle{\vphantom1\smash{\lower.5ex\hbox{\small$#1$}}%
	\over\vphantom1\smash{\raise.25ex\hbox{\small$#2$}}}}}

\def\p#1{\mskip#1mu}
\def\n#1{\mskip-#1mu}
\def\stop{\p6.}
\def\comma{\p6,}

\def\noj#1,#2,{{\bf #1} (19#2)\ }
\def\jou#1,#2,#3,{{\sl #1\/ }{\bf #2} (19#3)\ }
\def\ann#1,#2,{{\sl Ann.\ Physics\/ }{\bf #1} (19#2)\ }
\def\cmp#1,#2,{{\sl Comm.\ Math.\ Phys.\/ }{\bf #1} (19#2)\ }
\def\cq#1,#2,{{\sl Class.\ Quantum Grav.\/ }{\bf #1} (19#2)\ }
\def\cqg#1,#2,{{\sl Class.\ Quantum Grav.\/ }{\bf #1} (19#2)\ }
\def\ijmp#1,#2,{{\sl Int.\ J.\ Mod.\ Phys.\/ }{\bf A#1} (19#2)\ }
\def\jmp#1,#2,{{\sl J.\ Math.\ Phys.\/ }{\bf #1} (19#2)\ }
\def\grg#1,#2,{{\sl Gen.\ Rel.\ Grav.\/ }{\bf #1} (19#2)\ }
\def\mpl#1,#2,{{\sl Mod.\ Phys.\ Lett.\/ }{\bf A#1} (19#2)\ }
\def\nc#1,#2,{{\sl Nuovo Cim.\/ }{\bf #1} (19#2)\ }
\def\np#1,#2,{{\sl Nucl.\ Phys.\/ }{\bf B#1} (19#2)\ }
\def\pl#1,#2,{{\sl Phys.\ Lett.\/ }{\bf #1B} (19#2)\ }
\def\pla#1,#2,{{\sl Phys.\ Lett.\/ }{\bf #1A} (19#2)\ }
\def\pr#1,#2,{{\sl Phys.\ Rev.\/ }{\bf #1} (19#2)\ }
\def\prd#1,#2,{{\sl Phys.\ Rev.\/ }{\bf D#1} (19#2)\ }
\def\prl#1,#2,{{\sl Phys.\ Rev.\ Lett.\/ }{\bf #1} (19#2)\ }
\def\prp#1,#2,{{\sl Phys.\ Rept.\/ }{\bf #1C} (19#2)\ }
\def\ptp#1,#2,{{\sl Prog.\ Theor.\ Phys.\/ }{\bf #1} (19#2)\ }
\def\ptpsup#1,#2,{{\sl Prog.\ Theor.\ Phys.\/ Suppl.\/ }{\bf #1} (19#2)\ }
\def\rmp#1,#2,{{\sl Rev.\ Mod.\ Phys.\/ }{\bf #1} (19#2)\ }
\def\yadfiz#1,#2,#3[#4,#5]{{\sl Yad.\ Fiz.\/ }{\bf #1} (19#2) #3%
\ [{\sl Sov.\ J.\ Nucl.\ Phys.\/ }{\bf #4} (19#2) #5]}
\def\zh#1,#2,#3[#4,#5]{{\sl Zh.\ Exp.\ Theor.\ Fiz.\/ }{\bf #1} (19#2) #3%
\ [{\sl Sov.\ Phys.\ JETP\/ }{\bf #4} (19#2) #5]}

\def\lae{\mathrel{\mathop{\smash{\lower .5 ex \hbox{$\stackrel<\sim$}}}}}
\def\lae{\mathrel{\mathop{\smash{\lower .5 ex \hbox{$\stackrel>\sim$}}}}}

\def\pa{\partial}

\def\l:{\mathopen{:}\,}
\def\r:{\,\mathclose{:}}

\nblack

\catcode`\@=11
\def\theequation{\thesection.\arabic{equation}}
\@addtoreset{equation}{section}
\@addtoreset{footnote}{section}
\@addtoreset{footnote}{subsection}
\def\appendix{{\newpage\section*{Appendices}}\let\appendix\section%
{\setcounter{section}{0}
\gdef\thesection{\Alph{section}}}\section}
\catcode`\@=12

\begin{document}

\begin{titlepage}

\begin{center}
\today\hfill    TAUP-2170-94 \\
\hfill    hep-th/yymmddd

\vskip 1 cm

{\large \bf
$(1,q=-1)$ Model as a Topological Description of $2d$ String Theory}

\vskip 1.8 cm

{
	  Yoav Lavi, Yaron Oz {\it and\/} Jacob Sonnenschein\footnote{
Work supported in part by the US-Israel Binational Science Foundation,
and the Israel Academy of Science.\\e-mail:
yoavl@taunivm.bitnet, yarono@ccsg.tau.ac.il, cobi@taunivm.bitnet.}
}
\vskip 0.2cm

{\sl
School of Physics and Astronomy\\Raymond and Beverly Sackler Faculty
of Exact Sciences\\Tel-Aviv University\\Ramat Aviv, Tel-Aviv 69978, ISRAEL.
}

\end{center}

\vskip 1 cm

\begin{abstract}

We study the $(1,q=-1)$ model coupled to topological
gravity as a candidate to describing $2d$ string theory at the
self-dual radius.
We define the model by analytical continuation of
$q>1$ topological recursion relations to $q=-1$.
We show that at genus zero the $q=-1$ recursion relations yield
the $W_{1+\infty}$ Ward identities for tachyon correlators
on the sphere.
A scheme for computing
correlation functions of $q=-1$ gravitational descendants
is proposed and applied for the computation of several correlators. It is
suggested that the latter correspond to correlators of discrete states of the
$c=1$ string.
In a similar manner to the $q>1$ models, we show that
there exist topological recursion relations for the correlators
in the $q=-1$ theory that consist of only one and two splittings
of the Riemann surface.
Using a postulated regularized contact, we prove that
the genus one $q=-1$ recursion relations for tachyon correlators coincide
with the $W_{1+\infty}$ Ward identities on the torus.
We argue that the structure of these recursion relations
coincides with that of the $W_{1+\infty}$ Ward identities for any genus.

\end{abstract}
\newpage

\end{titlepage}

\section{Introduction}

Non-critical string theory in general and the $c\leq 1$ string models in
particular
attracted much attention in recent years. They provide
a framework for studying fundamental questions of string theory
and quantum gravity such as non-perturbative structure.
The most challenging of these class of models is the
$c=1$ string.
It differs from the $c<1$ string models by that it has a space-time
description and a propagating massless degree of freedom, the
tachyon. Furthermore,
BRST analysis in the continuum description of the theory
revealed the existence of special states at discrete values of momentum
\cite{Lian,Mukherji,Bouk,Wittengr,KPol},
one of which was argued to correspond to a discrete graviton describing
the two-dimensional black hole of \cite{Wittenbh}.

$c<1$ non-critical string models constructed by
coupling $c<1$ conformal matter to two-dimensional
gravity, have been studied using a variety of
methods reflecting their integrable and topological structures.
Their topological phase is realized by
$(1,q)$ minimal topological matter coupled to topological
gravity \cite{KekeLi,KekeLi2}, from which they can be reached by appropriate
perturbations. Intersection theory interpretation of the correlators
has been given in \cite{Wittalgeom,WittenNM}.

A $q^{th}$ KdV integrable hierarchy underlies the $(1,q)$
topological models and the corresponding $(p,q)$
non-critical strings \cite{Douglas}. This together with the string equation
provides a complete description of the theories.
The generating functions for correlators of the theories are
$\tau-$functions of the corresponding integrable hierarchies
\cite{dvvl,fukuma}.
Matrix integral representations of the generating functions have been
suggested by \cite{Konts,Miron,Adler}.
For the case $q=2$, that is
pure topological gravity, they reduce to
the Kontsevich integral \cite{Konts}.
However, while the latter integral has a nice interpretation in terms
of cell decomposition of the moduli space of Riemann surfaces, the
geometrical meanings of the general integrals are less clear.
Nevertheless, these matrix integrals provide a link between the
integrable and the topological structures.
A manifestation of the integrable and topological
structures is also provided by a set of Ward
identities given by $W_q$ constraints on the partition function
of the theory. This is argued to be equivalent to specifying the
integrable hierarchy together with a string equation \cite{Adler}
and to a set of topological recursion relations \cite{MR} as we discuss
in the sequel.

Another relation between the integrable and the topological structures
of the theories has been discovered using the topological Landau-Ginzburg
formulation \cite{DVV}. The integrable hierarchy appears in the
Landau-Ginzburg description in its Lax formulation. The
Landau-Ginzburg superpotential is identified with the Lax operator
in the dispersionless limit.

There exist two types of topological recursion relations for correlators
of $(1,q)$ theories. The basic strategy behind both is
analysis of contributions to the correlators from the boundary of
the moduli space.
One set of recursion relations has been derived, first for pure
topological gravity in \cite{WittTG,Wittinter} and later
generalized to topological matter
coupled to topological gravity \cite{WittenNM}.
The second set has been derived for pure topological gravity in
\cite{VV}. In the latter the concept
of contact algebra was introduced, which together with the requirement of the
 invariance
of the correlators under the interchange of operators provides
a complete solution of the theory.
The, thus derived, topological recursion relations coincide with the Virasoro
constraints on the partition function of the theory.
The uniqueness of the contact algebra underlying the theory has
been proved in \cite{AMS}.

The generalization of the contact algebra to $(1,q)$ models coupled
to topological gravity was suggested in \cite{MR},
and was argued to correspond to the $W_q$ constraints on the partition
function.
A key role in this topological procedure is played by
multicontacts whose importance was realized in \cite{KekeLi2,dvvl,AMS}.
Correlators of general $(1,q)$ models on the sphere, as well as
those of $(1,3)$ model on higher genus, were consistently calculated using the
topological scheme. The computation of genus $g$ correlators for
$q>3$ requires a consistent regularization scheme which is still
lacking. However, as shown in \cite{LaSo}, one can overcome this
problem since the recursion relations can be recast in a form that
involves only one and two contacts.

The topological description of the $c=1$ string is conjectured
to correspond to $(1,q)$ model coupled to topological gravity,
analytically continued to $q=-1$ or equivalently, to
$N=2$ twisted minimal model coupled to topological gravity.
Indeed the cohomology of the latter, realized an $SU(2)/U(1)$ coset at
level $k=-3$, coincides with that of the $c=1$ string \cite{MV,G/G}.
Interpretation of intersection theory calculations
of correlators at $k=-3$ as correlators in $c=1$ string theory is
in agreement with matrix model results.
First, the partition functions were observed to be identical
\cite{WittenNM}.
Second, the four-point correlator computed in \cite{WittenNM}, when
analytically to $k=-3$ was shown to be the tachyons four-point function
\cite{MV}.
Third, $1\rightarrow n$ amplitudes as well as five-point function at
various kinematic regions agree \cite{DMPint}.

The integrable hierarchy underlying the $c=1$ string is the Toda-lattice
hierarchy \cite{DMP}. The generating function for tachyon correlators
is a $\tau-$function of the hierarchy. There exist a matrix integral
representation of the latter \cite{DMP},
generalizing that of the $c<1$ case
\footnote{A two-matrix integral representation of the generating
function for tachyon correlators has been proposed in \cite{Bonora}.}.
Recently the string equations were constructed \cite{Tak,Eguchi}.

The topological Landau-Ginzburg description of the theory is
constructed as the $A_{k+1}$ model at $k=-3$ \cite{HOP,GM}.
A relation between the integrable and the topological structures is
established by that the Landau-Ginzburg superpotential is the
Baker-Akhiezer wave function of the Toda lattice hierarchy \cite{HOP}.
This parallels the identification of the superpotential with the
Lax operator in the $c<1$ cases.
There exist $W_{1+\infty}$ Ward identities for tachyon correlators
in the $c=1$ string \cite{MP,DMP}. These relations determine completely
the tachyon dynamics. The $W_{1+\infty}$ algebra of constraints
on the partition function is probably the $c=1$ analog of the
$W_q$ algebra of constraints in the $c<1$ cases.
These Ward identities can be derived from the Toda lattice
and the string equations \cite{Tak,Eguchi}. On the other hand they coincide
with period integrals in the topological Landau-Ginzburg formulation
of the theory \cite{HOP}. These provide another link between the topological
and the integrable structures of the $c=1$ string.

As described, the integrable and topological structures of the $c=1$ case
parallel those of the $c<1$ cases. One of the main differences, however,
is the lack of topological recursion relations in the former.
It is straightforward to verify that Witten's topological recursion
relations are no longer correct for the $c=1$ string, and should
be modified. However, the required modification is still unknown.
Our aim in this paper is to generalize the second set of topological
recursion relations as proposed in \cite{MR}
for the $c<1$ models, and provide a topological procedure for
computation of correlators in $c=1$.
The idea will be to analytically continue the $(1,q)$
recursion relations to $q=-1$.
We will show explicitly for genus zero
and one that the analytically continued topological recursion
relations coincide with the $W_{1+\infty}$ Ward identities for
tachyon correlators, and will argue that their structures coincide
for any genus.
We will further use the topological recursion relations to compute
correlators of $q=-1$ gravitational descendants which are suggested
to correspond to discrete states of the $c=1$ string.
Similarly to the $c<1$ cases, we show that the topological recursion
relations can be recast in a form that involves only one and two splittings
of the Riemann surface. This provides us with another set of Ward identities
for the theory.

The paper is organized as follows. In section two we review the
$(1,q)$ topological models and the topological
procedure to compute their correlators.
In section three the $W_{1+\infty}$ Ward identities of $2d$ string theory
at the self-dual radius are described.
In section four we pose the rules for the analytic continuation of the
topological recursion relations from $q>1$ to $q=-1$.
We use the analytically continued scheme
to compute up to five point tachyon correlators at genus zero.
The results are in agreement with $c=1$ matrix model calculations.
The equivalence  between the
$(1,q=-1)$ recursion relations and the $W_{1+\infty}$ Ward identities
at genus zero is proven in section five.
In section six we propose a topological
scheme for computation of correlators of $q=-1$ gravitational descendants,
which we interpret as discrete states and compute
several correlators.
Topological recursion relations in terms of one and two splittings
are written in section seven.
Section eight is devoted to the study of higher genus recursion relations.
We define a regularized contact and use it to prove the equivalence
between the recursion relations and the $W_{1+\infty}$ Ward identities
on the torus. Two and three point tachyon correlators on the torus are
computed via the topological procedure.
Section nine is devoted to discussion and conclusions.
In appendix A we include an example of computation of three-point
tachyon correlator on the sphere as well as two-point
tachyon correlator on the torus
using the new recursion relations introduced in
section seven.

\section{$(1,q)$ topological models}
$(1,q)$ models form a special sub sector of the $(p,q)$ minimal models.
Having zero physical fields, they are not well defined conformal field
theories. However, they make sense as topological field theories, the
so called topological minimal models \cite{KekeLi,KekeLi2}.
The observables consist of $q-1$
primary fields $P_{0,\alpha} , \alpha = 1,..,q-1$.
When coupled to topological gravity, a family of gravitational descendants
$P_{k,\alpha}$ is associated with each primary field $P_{0,\alpha}$,
where $k$ takes positive
integer values. For $q=2$ we get topological gravity with
$P_{0,1}$ as the puncture operator. $P_{k,1}$ correspond in this case
to the Mumford, Morita and Miller
cohomology classes on the compactified moduli space
$\bar{{\cal M}}_{g,s}$.

An integer ghost number is attributed to the fields:
\beq
gh(P_{k,\alpha}) = (k-1)q + (\alpha - 1)\stop
\label{gh}
\eeq
{}From the viewpoint of the integrable structure
the ghost number basically corresponds to the power of the KP Lax operator
associated with the field \cite{DMR}
, while from the viewpoint of the topological
structure it corresponds to the degree of the form on the moduli
space associated with it \cite{Wittalgeom,WittenNM}.

The ghost number conservation law for the genus $g$ correlator
$\langle \prod_{i=1}^sP_{k_i,\alpha_i}\rangle _g$ reads
\beq
\sum_{i=1}^s gh(P_{k_i,\alpha_i}) = 2(g-1)(1+q)\stop
\label{colaw}
\eeq
The ghost number conservation follows from the requirement
of having a residue in the KdV computational scheme,
and from the demand that the total form associated with the
correlator be a top form on the moduli space in the
corresponding intersection theory.

An equivalent description, which is naturally analytically continued to the
$q=-1$ case, is the following:
$P_{k,\alpha} \rightarrow P_n$ where $n = kq + \alpha$. $P_n$ are in $1-1$
correspondence with $P_{k,\alpha}$ :
\beq
\alpha = n~ mod~ q,~~~~~~~~ k = \frac{(n - \alpha)}{q}.
\eeq
The conservation law for the correlator
$\langle \prod_{i=1}^s P_{n_i}\rangle _g$
is
\beq
\sum_{1=1}^s n_i = (s + 2g -2)(q+1)\stop
\label{cgh}
\eeq

A topological procedure to calculate
correlation functions of the model which is equivalent to the $W_q$
constraints has been developed in \cite{MR}.
In contrast to the $W_q$ Ward identities which are complicated and
are not known in general, the topological procedure consists of
simple topological rules.
The idea behind the scheme is that the
correlators can be determined by contacts between the operators and
between them and the degenerations of the Riemann surface.
This procedure yields topological recursion relations generalizing
those proposed by \cite{VV} for topological gravity, i.e. the
$(1,2)$ model.
Let us briefly review it.

The first thing to notice is that the metric on the space of physical fields,
defined by the genus zero two-point function, vanishes, since
\beq
\eta_{ij} \equiv \langle P_iP_j\rangle _0 = |i|\delta_{i+j,0} \comma
\label{metric}
\eeq
while the physical fields of the theory are $P_n$ with $n$ being  positive
integer charges.
In order to overcome this difficulty, auxiliary unphysical
fields with negative charges are introduced.
These fields appear in the metric and
decouple from higher point functions.
Note that the definition of the metric (\ref{metric})
is differs from the standard one \cite{DW}:
$\eta_{i,j}=\langle P_1P_iP_j\rangle _0$ with $P_1$ being the puncture operator
 and
$P_i,P_j$ are primary operators. The metric (\ref{metric}) is defined on
the space of both primaries and descendants.

An identity operator to be inserted in degenerations is constructed in the
usual way:
\beq
I = \sum_{i,j}|P_i\left>\eta^{ij}\right<P_j| \stop
\label{id}
\eeq

As a consequence of introducing negative charge fields there exists a
one point
function that does not vanish on the sphere
\beq
\langle P_{-q-1}\rangle _0 = -q \stop
\label{1pts}
\eeq

Consider now the genus $g$ correlation function
$\langle P_n\prod_{i=1}^m P_{n_i}\rangle _g$.
Denote $P_n$ as the marked operator, that is the operator that performs
contacts in this procedure. It has contacts with
$\alpha = n~ mod~ q$ operators.
The contact algebra reads:
\beq
\overbrace{P_nP_{i_1}...P_{i_{\alpha}}} = P_{n +  \sum_{k=1}^{\alpha}
i_k - \alpha(q+1)}\comma
\label{cont}
\eeq
where over brace means contact.
Contacts and degenerations are the ingredients
for computing correlators in this scheme.
At each degeneration one inserts a complete set
of states.
The topological procedure of \cite{MR} is summarized
by the degeneration equation:
\beq
\sum_{\Delta} \langle P_n\prod_{i = 1}^m P_i\rangle  = 0 \comma
\label{degeq}
\eeq
where $\sum_{\Delta}$ means summation over all the degenerations with the first
operator, i.e. $P_n$, performing the contacts.
The contributions to the degeneration equation come from the boundary
of the moduli space and are of three types: splitting, pinching of
dividing cycles and pinching of nontrivial homology cycles.

The degeneration equation yields Ward identities for the $(1,q)$ models
coupled to topological gravity.
As an example to that consider the $(1,2)$ model.
The degeneration equation for  $\langle P_n\prod_{i=1}^m P_{n_i}\rangle _g$
 reads
\beqar
&&\sum_j\langle \overbrace{P_nP_j}\prod_{i=1}^m P_{n_i}\rangle _g\langle
 P_{-j}\rangle _0 +
\sum_{j;k=1}^m \langle \overbrace{P_nP_j}\prod_{k \neq i=1}^m
P_{n_i}\rangle _g\langle P_{-j}P_{n_k}\rangle _0 + \nonumber\\&&
\sum_{g';j}\langle \overbrace{P_nP_j}\prod_{k\in S_1}
P_{n_k}\rangle _{g'}\langle P_{-j}\prod_{l \in S_2} P_{n_l}\rangle _{g-g'} +
 \nonumber\\
&&\sum_j\langle \overbrace{P_nP_j}P_{-j}\prod_{i=1}^m P_{n_i}\rangle _{g-1} = 0
\comma
\label{VVe}
\eeqar
with $S_1 \cup S_2 = (1...m)$.

As shown in Fig.1, the first two terms in (\ref{VVe}) correspond to
splitting, the third to pinching of a dividing cycle
and the last term corresponds to pinching a nontrivial homology cycle.
Note that the second term in (\ref{VVe})
may be considered from the degeneration equation
viewpoint as a special case of the third term.


\setlength{\unitlength}{1mm}

\newcommand\putfig[3]{
   \vbox{
   \let\picnaturalsize=N
   \def\picsize{#3}
   \def\picfilename{#1}
   \ifx\nopictures Y\else{\ifx\epsfloaded Y\else\input epsf \fi
   \let\epsfloaded=Y
   \centerline{\ifx\picnaturalsize N\epsfxsize \picsize\fi
   \epsfbox{\picfilename}}}\fi
   \vspace{1.0cm}
   {\it #2}
   \vspace{1.5cm}
   }
}

\vskip 0.5cm
\putfig{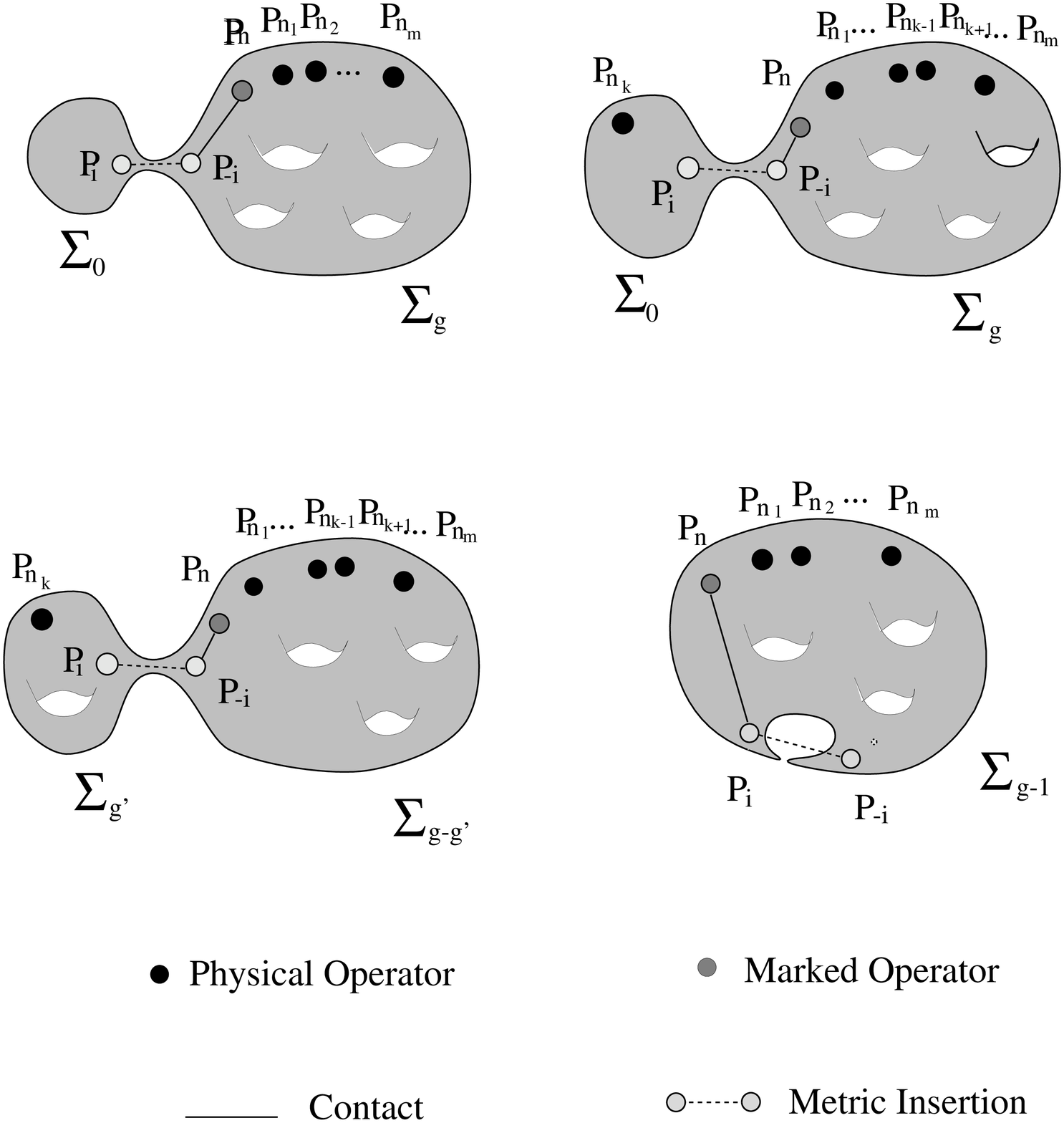}{ Fig. 1: The degeneration equation for
topological gravity}{100mm}

Using (\ref{metric}), (\ref{1pts}) and (\ref{cont}) the degeneration
equation (\ref{VVe}) takes the form
\beqar
\langle P_n\prod_{i=1}^m P_{n_i}\rangle _g = \frac{1}{2}[
\sum_{j=1}^m n_j\langle P_{n+n_j-3}\prod_{j\neq i=1}^m
P_{n_i}\rangle _g + \nonumber\\
\sum_{g';j}\langle P_{n+j-3}\prod_{k \in S_1}
P_{n_k}\rangle _{g'}\langle P_{-j}\prod_{l \in S_2} P_{n_l}\rangle _{g-g'} +
 \nonumber\\
\sum_j\langle P_{n+j-3}P_{-j}\prod_{i=1}^m P_{n_i}\rangle _{g-1}]
\comma
\label{VVrr}
\eeqar
where we require the correlators with auxiliary fields to vanish.
Equation (\ref{VVrr}) is the Verlinde's recursion relation for
topological gravity
\cite{VV}, with a
difference of notation due to the difference between the definition
of the ghost number (\ref{gh}) and the degree of the Mumford-Morita-Miller
cohomology classes. It should be stressed, however,
that we derive the recursion relations via a different procedure than that
of \cite{VV}.

Our aim is to relate the topological phase of $2d$ string theory
to the $(1,q=-1)$ model coupled to topological gravity, defined
via analytical continuation of the $q>1$ degeneration equation to $q=-1$.
This will provide us with topological recursion relations,
which, as we shall argue, reduce for tachyon correlators to
the $W_{1+\infty}$ Ward identities of $2d$ string theory
at the self-dual radius.

\section{$W_{1+\infty}$ Ward identities of $2d$ string theory}

Tachyon dynamics in $2d$ string theory has been studied in the
continuum \cite{kdf} as well as in the matrix formulation of the
theory \cite{twodreview}. The full scattering matrix was computed \cite{MPR}
and a set of $W_{1+\infty}$ constraints on the amplitudes was
derived \cite{DMP,MP}. The latter form Ward identities that determine
the tachyon correlators completely.
In this section we review these $W_{1+\infty}$ Ward identities and
derive various formulas that will be needed later.

Introduce the notation:
\beq
\langle O\rangle   \equiv \sum_{g\geq 0} \frac{1}{\mu^{2g-2}} \langle O\rangle
 _g \comma
\eeq
where expansion in $\frac{1}{\mu^2}$ corresponds to genus expansion,
and $\langle O\rangle _g$ is the genus $g$ correlator of $O$.

The $W_{1+\infty}$ Ward identities of $2d$ string theory read
\cite{DMP}:

\beq
\langle \langle T_n\rangle \rangle  \equiv \langle T_n
 exp[\sum_{k=-\infty}^{\infty}t_kT_k]\rangle  =
\bar{W}^{(n+1)}_{-n} Z \comma
\label{Wcons}
\eeq
where
\beq
Z \equiv \langle exp[\sum_{k=-\infty}^{\infty}t_kT_k]\rangle  \stop
\label{Z}
\eeq
$T_n$ is the tachyon of integer momentum $n$, and $t_n$ is the time
associated with it.
$\bar{W}^{(n+1)}_{-n}$ is the $-n$ mode of a spin $n+1$ current
$\bar{W}^{(n+1)}(x)$ and is given by \cite{DMP}
\beq
\bar{W}^{(n+1)}_{-n} = \oint {\rm d}x \frac{(i\mu)^{-(n+1)}}{n+1}
:e^{-i\mu\varphi(x)}\partial_x^{n+1}e^{i\mu\varphi(x)}:
\comma
\label{Wcur}
\eeq
where
\beq
\partial\varphi(x) =\frac{1}{x}[1+ \sum_{k>0}t_{-k}x^k + \frac{1}{\mu^2}
\sum_{k>0}kx^{-k}\partial_{-k}] \comma
\label{phi}
\eeq
with $\partial_{-k} \equiv \frac{\pa}{\pa t_{-k}}$.

Evaluating (\ref{Wcons}) we get
\beqar
\langle \langle T_n\rangle \rangle _0 & = &
\frac{1}{n(n+1)} res(\bar{W}^{n+1}_0)  \comma \nonumber\\
\langle \langle T_n\rangle \rangle _1 & = & \frac{1}{n}
 res(\bar{W}^{n}_0\bar{W}_1)
-\frac{1}{24}(n-1)
res(\bar{W}^{n-2}_0\bar{W}''_0)  \comma \nonumber\\
\langle \langle T_n\rangle \rangle _g & = & \frac{1}{n}
 res(\bar{W}^{n}_0\bar{W}_g +
n\bar{W}^{n-1}_0\bar{W}_1 \bar{W}_{g-1} +... ) ~~~~~g > 1
\comma
\label{WIw}
\eeqar
where $res$ means picking the $x^{-1}$ term in the Laurant expansion,
$prime$ denotes a derivative with respect to $x$ and
\beqar
\bar{W}_0 & = & \frac{1}{x}[1+ \sum_{k>0}t_{-k}x^k +
\sum_{k>0}kx^{-k}\langle \langle T_{-k}\rangle \rangle _0 ] \comma
\nonumber\\
\bar{W}_g & = & \sum_{k>0}kx^{-k-1}\langle \langle T_{-k}\rangle \rangle _g
 ~~~~~g \geq 1 \stop
\label{Wexp}
\eeqar

Define $\Phi_n^{(g)} = \pa_n \bar{W}_g$, then
\beqar
\Phi_n^{(0)} & = &\Theta(-n)x^{-n-1} + \sum_{k>0}x^{-k-1}\langle \langle
 T_nT_{-k}\rangle \rangle _0 \comma
\nonumber\\
\Phi_n^{(g)} & = & \sum_{k>0}x^{-k-1}\langle \langle T_nT_{-k}\rangle \rangle
_g
 \stop
\label{Phi}
\eeqar
In the topological Landau-Ginzburg formulation of the theory
$\bar{W}_0$ and $\Phi_n^{(0)}$
correspond to the superpotential and to the Landau-Ginzburg
field respectively \cite{HOP}.

For the explicit expansion of the $W_{1+\infty}$ we will need the
following formula:
\beq
\partial_{n_1}...\partial_{n_{m-1}}\Phi_{n_m}^{(g)}(t = 0) =
(\sum_{i = 1}^m n_i) \Theta(\sum_{i = 1}^m n_i)
x^{-1-\sum_{i = 1}^m n_i}
\langle T_{- \sum_{i = 1}^m n_i}\prod_{i = 1}^m T_{n_i}\rangle _g\stop
\label{part}
\eeq

In the sequel we will also need:
\beqar
&&\pa_{-n}\bar{W}_0^{''}(t=0)  = (n-1)(n-2)x^{n-3} \comma \nonumber\\
&&\partial_{n_1}...\partial_{n_m}\bar{W}_g^{''}(t = 0)  =
(\sum_{i = 1}^m n_i)(1+ \sum_{i = 1}^m n_i)(2+\sum_{i = 1}^m n_i)
\nonumber\\
&&\Theta(\sum_{i = 1}^m n_i)x^{-3-\sum_{i = 1}^m n_i}
\langle T_{- \sum_{i = 1}^m n_i}\prod_{i = 1}^m T_{n_i}\rangle _g~~~~~~~g \geq
 0\stop
\label{Wpart}
\eeqar

\section{Genus zero tachyon correlators via $(1,q=-1)$ theory }

The approach that we take in order to define the $(1,q=-1)$ theory
is to analytically continue the $q>1$ degeneration equation.
This will provide us with a set of topological recursion relations
for $2d$ string theory at the self-dual radius.
The rules for the analytical continuation are the following:
(i) we consider correlators of physical operators in $q>1$ models,
that is $P_n$ with $n$ positive and allow negative values only at the
final analytically continued recursion relations.
(ii) The argument of the Heaviside function $\Theta(x)$
that appears in the $q>1$ degeneration equation,
due to the decoupling of the auxiliary fields, will change sign
at $q=-1$. The reasoning for this will be given in the sequel.

In order to demonstrate the analytic continuation procedure
let us compute the tachyon correlators up to
the five point function on the sphere
\footnote{From now on, unless explicitly stated, we consider genus
zero correlators.}.


Consider first the two-point function
\beq
\langle P_nP_{-n}\rangle  = n \comma
\label{2pt}
\eeq
where $P_n$ being a primary operator.
At $q=-1$ we identify the primary operator $\frac{P_n}{n}$ as the
positive momentum tachyon $T_n$
and the auxiliary operator $\frac{P_{-n}}{-n}$ as the negative momentum
tachyon $T_{-n}$, thus
\beq
\langle T_nT_{-n}\rangle  = -\frac{1}{n} \stop
\label{t2pt}
\eeq
Equation (\ref{t2pt}) differs by sign from the conventions
of $2d$ string matrix model \cite{DMP}. This overall sign difference will
 persist
for all the tachyons correlators.

We have not used the degeneration equation yet, but it is used  already
for computing the tachyons three point function.
We assume in the following that the marked operator is primary.
In section 6 we will consider the case when the marked operator is
a gravitational descendant.
Consider the correlator $\langle P_nP_{n_1}P_{n_2}\rangle $,
with all the operators in the correlator being physical.
Thus in the framework of the $(1,q)$ models,
the charges $n_i$ are positive, and we allow negative values
in the final analytically continued formula.

Taking $P_n$ as the marked operator, the degeneration equation reads:
\beq
(n+1)(-q)^n \langle P_nP_{n_1}P_{n_2}\rangle  + (n+1)n(-q)^{n-1}\langle
 P_{-n_1}P_{n_1}\rangle \langle P_{-n_2}
P_{n_2}\rangle  = 0 \stop
\label{3pte}
\eeq
with the different terms depicted in Fig. 2.

\vskip 0.5cm
\putfig{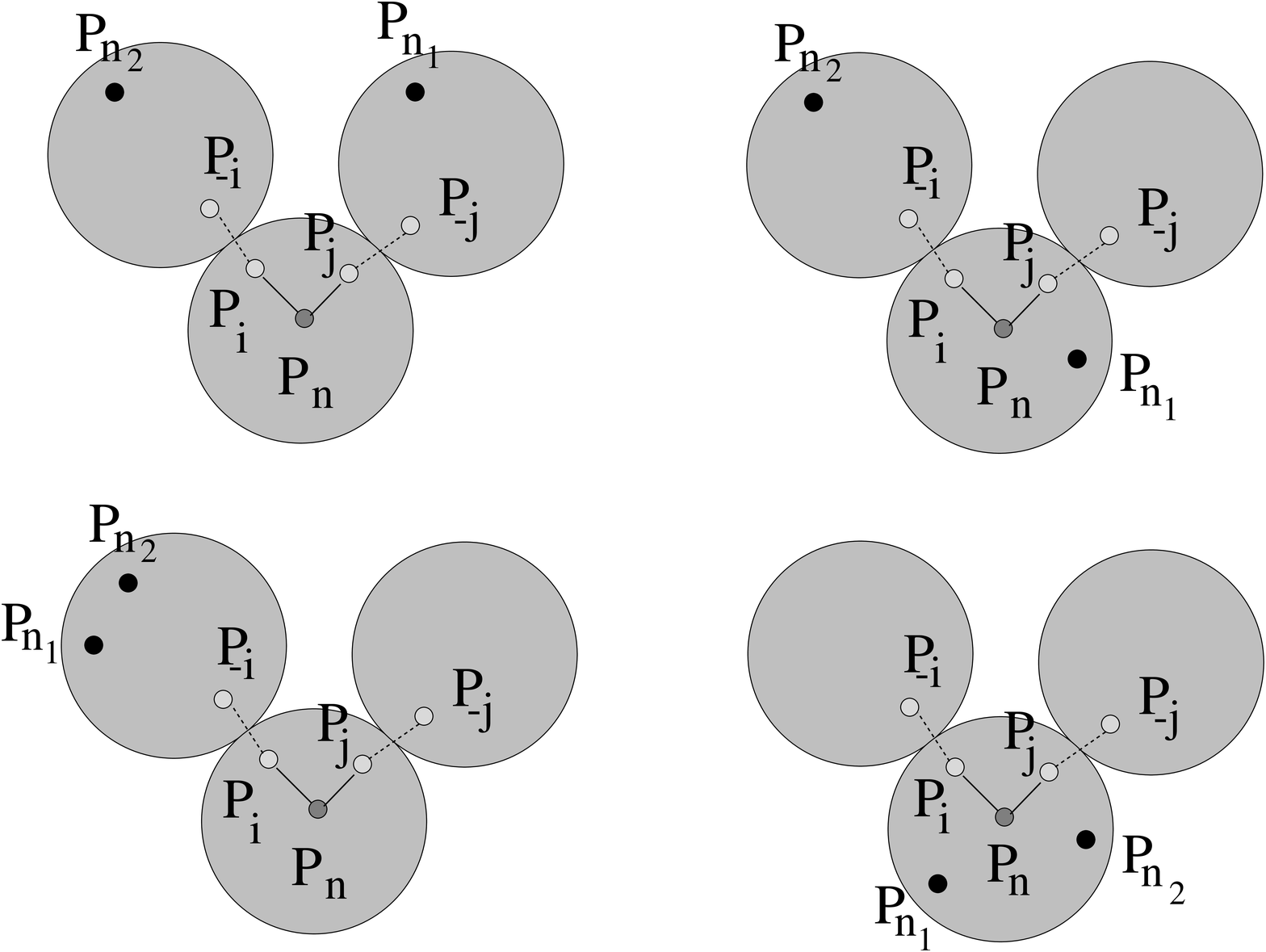}{ Fig. 2: The degeneration equation for three-point
function on the sphere}{100mm}

Solving for the required three-point function we get:
\beq
\langle P_nP_{n_1}P_{n_2}\rangle  =  \frac{nn_1n_2}{q} \stop
\label{3pt}
\eeq
After normalizing we have at $q=-1$
\beq
\langle T_nT_{n_1}T_{n_2}\rangle  =  -1 \stop
\label{T3pt}
\eeq
Note that we can get the same answer by taking the marked operator to
be $P_{-n}$ and formally performing $-n$ contacts. This parity invariance
property of the procedure implies that we may think of $T_{-n}$ as
primaries with negative charges. We will further
 discuss this issue in section 6.

Consider now the correlator $\langle P_nP_{n_1}P_{n_2}P_{n_3}\rangle $.
The degeneration equation reads
\beqar
(-q)^n \langle P_nP_{n_1}P_{n_2}P_{n_3}\rangle  +  n(-q)^{n-1}\sum_{i = 1}^3
\langle P_{n + n_i -(q+1)}\prod_{i \neq j = 1}^3P_{n_j}\rangle \nonumber\\
\langle P_{-n_i}P_{n_i}\rangle
+ n(n-1)(-q)^{n-2}\prod_{i = 1}^3\langle P_{-n_i}P_{n_i}\rangle  = 0\comma
\label{e4pt}
\eeqar
where we divided by the overall prefactor $n+1$.
Consider the term $\langle P_{n + n_i-(q+1)}\prod_{i \neq j =
1}^3P_{n_j}\rangle
 $ in
(\ref{e4pt}).
In the $(1,q)$ models with $q>1$ such a correlator is proportional to
$\Theta(n + n_i - (q+1))$ since auxiliary fields should decouple from
all the correlators besides the two-point functions. In a formulation that
is more adequate for the $W_q$ constraints approach this
can also be stated as  $\Theta((k + k_i-1)+
{(\alpha+\alpha_i -1)\over q})$.   Thus, when we
 analytically
continue to $q = - 1$
the correlator becomes proportional to
$\Theta(-n - n_i)$. This $\Theta$ term implies that there is no contact
between two positive momentum tachyons as has been found in the topological
Landau-Ginzburg description of the $c=1$ string \cite{HOP,GM}.

Taking the fields in (\ref{e4pt}) as primaries or their auxiliary analogs
and setting $q = -1$ we get the tachyons four point function
\beq
\langle T_nT_{n_1}T_{n_2}T_{n_3}\rangle  = - (n-1) + \sum_{i=1}^3 (n +
 n_i)\Theta(-n - n_i)
\stop
\label{t4pt}
\eeq

As a final example consider the five-point function
$\langle P_n\prod_{i=1}^4P_{n_i}\rangle $.
The degeneration equation reads
\beqar
&&(-q)^n\langle P_n\prod_{i=1}^4P_{n_i}\rangle  +
n(-q)^{n-1}\sum_{i = 1}^4
\langle P_{n + n_i - (q+1)}\prod_{i \neq j = 1}^4P_{n_j}\rangle \langle
 P_{-n_i}P_{n_i}\rangle
\nonumber\\
&&+n(-q)^{n-1}\sum_{i,j = 1, i \neq j}^4
\langle P_{n + n_i + n_j - 2(q+1)}\prod_{i,j \neq k = 1}^4P_{n_k}\rangle
\langle P_{-n_i - n_j +(q+1)}P_{n_i}P_{n_j}\rangle \nonumber\\
&&+n(n-1)(-q)^{n-2}\sum_{i,j = 1, i \neq j}^4
\langle P_{n + n_i + n_j - 2(q+1)}\prod_{i,j \neq k = 1}^4P_{n_k}\rangle
\langle
 P_{-n_i}P_{n_i}\rangle
\langle P_{-n_j}P_{n_j}\rangle \nonumber\\
&&+ n(n-1)(n-2)(-q)^{n-3}\prod_{i = 1}^4\langle P_{-n_i}P_{n_i}\rangle  = 0
 \stop
\label{5pt}
\eeqar

Following the same procedure as before we get at $q = -1$ the tachyons
five point function,
\beqar
\langle T_n\prod_{i=1}^4T_{n_i}\rangle  = - (n-1)(n-2) + \nonumber\\
(n-1)\sum_{i,j = 1, i \neq j}^4(n+n_i+n_j) \Theta(-n - n_i - n_j)
\langle T_{n + n_i + n_j}\prod_{i,j \neq k = 1}^4T_{n_k}\rangle     \nonumber\\
+ \sum_{i=1}^4 (n + n_i)
\Theta(-n - n_i)\langle T_{n + n_i}\prod_{i \neq j=1}^4T_{n_j}\rangle
\stop
\label{t5pt}
\eeqar
Note that the splitting to two three-point functions in (\ref{5pt}) vanishes
when we analytically continue to $q = -1$. This simplification of the
five point function will not persist for higher correlators.

The above computations describe the scheme for calculating tachyon
correlators
via the $q=-1$ model by analytical continuation of the degeneration equation.
Positive momentum tachyons are identified with primary operators, while
negative momentum tachyons are identified with their auxiliary analogs.
The latter decouple for $q > 1$ but not at $q=-1$. They provide us with
the negative times that are needed in order to pass from the KP integrable
hierarchy underlying the minimal models coupled to gravity to the Toda
lattice hierarchy underlying $2d$ string theory
\cite{DMP,HOP,Tak,Eguchi}.

\section{ The equivalence between the $(1,q=-1)$ degeneration equation
 and the $W_{1+ \infty}$ Ward identities for tachyon correlators
 at genus zero}

Our aim in this section is to prove that the genus zero
$(1,q=-1)$ degeneration equation
for tachyon correlators is identical to the genus zero
$W_{1+\infty}$
Ward identities for tachyon correlators in $2d$ string theory.

\subsection{ The $(1,q=-1)$ degeneration equation for genus zero
$n$ tachyons correlator}

Consider the genus zero
correlator $\langle P_n\prod_{i = 1}^m P_{n_i}\rangle $, where $P_n$
is a primary operator. Taking $P_n$ as the marked operator, the
degeneration equation reads:
\begin{eqnarray}
&&(-q)^n\langle P_n\prod_{i = 1}^m P_{n_i}\rangle  + n(-q)^{n-1}
\sum_{i = 1}^m \langle P_{n + n_i-(q+1)}\prod_{i \neq j = 1}^m P_{n_j}\rangle
\langle P_{-n_i}P_{n_i}\rangle + \nonumber\\
&&n(-q)^{n-1} \sum_{i,j = 1; i \neq j}^m
\langle P_{n + n_i + n_j-2(q+1)}\prod_{i,j \neq k = 1}^m P_{n_k}\rangle
\langle P_{-n_i -n_j +(q+1)}P_{n_i}P_{n_j}\rangle  + \nonumber\\
&&n(n-1)(-q)^{n-2} \sum_{i,j = 1; i \neq j}^m
\langle P_{n + n_i + n_j -2(q+1)}\prod_{i,j \neq k = 1}^m P_{n_k}\rangle
\langle P_{-n_i}P_{n_i}\rangle  \langle P_{-n_j}P_{n_j}\rangle  + \nonumber\\
&&... + \frac{\Gamma(n + 1)}{\Gamma(n - k + 1)}(-q)^{n-k}
\sum_{i_1..,i_k = 1; i_j \neq i_l}^m
\langle P_{n + n_{i_1} + ..+ n_{i_k}-k(q+1)}\prod_{i_1,..,i_k \neq i = 1}^m
 P_{n_i}\rangle
\prod_{l = 1}^k \langle P_{-n_{i_l}}P_{n_{i_l}}\rangle  \nonumber\\
&&+... + \frac{\Gamma(n + 1)}{\Gamma(n - m + 3)}(-q)^{n-m+2}
\sum_{j,k=1;j\neq k}^m
\langle P_{n + \sum_{j,k \neq i=1}^m n_i - (m-2)(q+1)}P_{n_j}P_{n_k}\rangle
\nonumber\\
&&\prod_{j,k\neq l=1}^m\langle P_{-n_l}P_{n_l}\rangle + ... +
\frac{\Gamma(n + 1)}{\Gamma(n - m + 2)}(-q)^{n-m}
\prod_{i = 1}^m \langle P_{-n_i}P_{n_i}\rangle  \stop
\label{qcorrel}
\end{eqnarray}

\vskip 0.5cm
\putfig{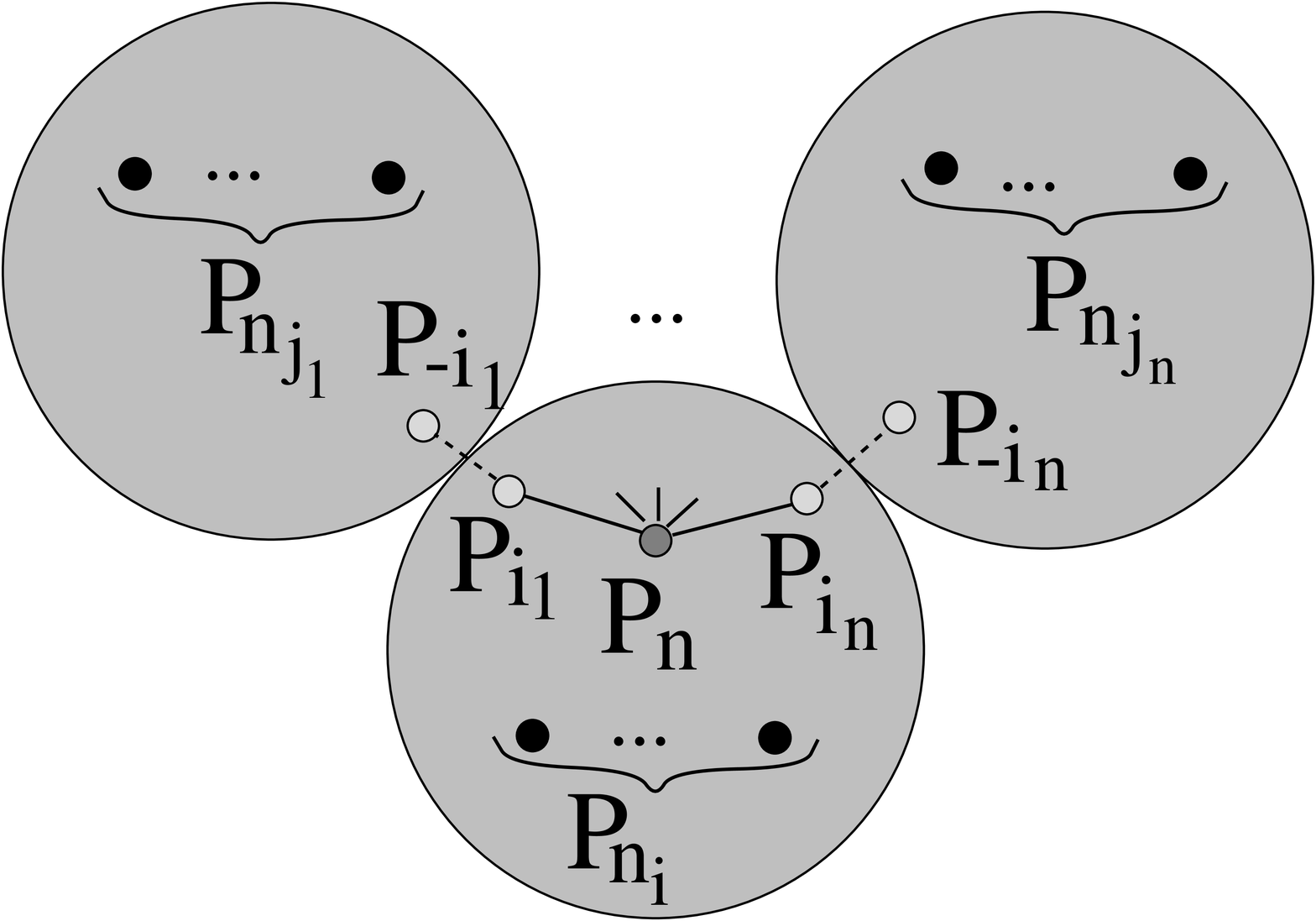}{ Fig. 3: A general term in the genus zero degeneration
equation}{100mm}

A general term in (\ref{qcorrel}) as depicted in Fig.3
is, up to a combinatorial factor,
of the form
\beq
\langle \overbrace{P_nP_{i_1}...P_{i_n}}\prod_{i \in S}P_{n_i}\rangle
\langle P_{-i_1}\prod_{j_1 \in S_1}P_{n_{j_1}}\rangle ...
\langle P_{-i_n}\prod_{j_n \in S_n}P_{n_{j_n}}\rangle  \comma
\label{gtp}
\eeq
where the sets $S,S_1..S_n$ are disjoint, possibly empty, and satisfy
$S \cup S_1..\cup S_n = (1...m)$. In order to derive (\ref{qcorrel})
we used (\ref{cont}).
Note that if $n$ is less than $m$ then some of the terms in equation
(\ref{qcorrel}) vanish.

Taking the operators to be primaries or their auxiliary analogs we
get at $q=-1$ a Ward identity for tachyon correlators:
\begin{eqnarray}
&&\langle T_n\prod_{i = 1}^m T_{n_i}\rangle
=  \sum_{i = 1}^m (n + n_i)\Theta(-n - n_i)
\langle T_{n + n_i}\prod_{i \neq j=1}^m T_{n_j}\rangle  +
\nonumber \\&&
\sum_{i,j = 1; i\neq j}^m (n + n_i + n_j)\Theta(-n - n_i - n_j)
(n_i + n_j)\Theta(n_i + n_j)
\langle T_{n + n_i + n_j}\prod_{i,j\neq k=1}^m T_{n_k}\rangle  +
\nonumber \\&&
(n-1) \sum_{i,j = 1;i\neq j}^m (n + n_i + n_j)\Theta(-n - n_i - n_j)
\langle T_{n + n_i + n_j}\prod_{i,j\neq k=1}^m T_{n_k}\rangle  +
...+\nonumber \\&&
\frac{\Gamma(n)}{\Gamma(n - k + 1)}
\sum_{i_1..,i_k = 1;i_j\neq i_l}^m
(n_{i_1}+..+ n_{i_k})\Theta(n_{i_1}+..+ n_{i_k})
\langle T_{n + n_{i_1} + ..+ n_{i_k}}
\prod_{i_1..i_k\neq j=1}^m T_{n_j}\rangle \nonumber \\&&
+...+ \frac{\Gamma(n)}{\Gamma(n - m + 3)}
\sum_{j,k = 1;j\neq k}^m (n + \sum_{j,k \neq i = 1}^m n_i)\Theta(-n -
\sum_{j,k \neq i = 1}^m n_i)
\langle T_{n + \sum_{j,k \neq i = 1}^m n_i }T_{n_j}T_{n_k}\rangle
\nonumber \\&&
+... - \frac{\Gamma(n)}{\Gamma(n - m + 2)}\stop
\label{tcorrel}
\end{eqnarray}
The last piece is recognized as the $1\rightarrow m$ amplitude and the
rest are contributions from other kinematic regions.
This recursion relations are highly non-linear
with a  general term consisting
of a product of tachyon correlators.

\subsection{$W_{1+ \infty}$ Ward identities for genus zero
$n$ tachyons correlator}

Consider the genus zero correlator $\langle T_{n}\prod_{i = 1}^m T_{n_i}\rangle
 $.
Using the Ward identities (\ref{WIw}) we have
\beqar
&&\langle T_{n}\prod_{i = 1}^m T_{n_i}\rangle  \equiv
 \partial_{n_1}...\partial_{n_m}
\langle \langle T_{n}\rangle \rangle (t=0) =
\frac{1}{n(n+1)} \partial_{n_1}...\partial_{n_m} res(\bar{W}_0)^{n+1} =
\nonumber\\&&
res[\frac{\Gamma(n)}{\Gamma(n - m + 2)}\Phi_{n_1}^{(0)}..
\Phi_{n_m}^{(0)} \bar{W}_0^{n - m + 1} + \nonumber\\&&
\frac{\Gamma(n)}{\Gamma(n - m +
 3)}(\partial_{n_1}(\Phi^{(0)}_{n_2}..\Phi^{(0)}_{n_m})
+ \Phi^{(0)}_{n_1}\partial_{n_2}(\Phi^{(0)}_{n_3}..\Phi^{(0)}_{n_m}) + .. +
 \Phi^{(0)}_{n_1}..
\partial_{n_{m-1}} \Phi^{(0)}_{n_m})\bar{W}_0^{n-m+2}  \nonumber\\&&
+...+
\frac{\Gamma(n)}{\Gamma(n - k + 1)}(\partial_{n_1}...\pa_{n_{m-k-1}}
(\Phi^{(0)}_{n_{m-k}}...\Phi^{(0)}_{n_m}) + ...)\bar{W}_0^{n-k} + \nonumber\\&&
+...+(n-1)(\partial_{n_1}..\partial_{n_{m-3}}(\Phi^{(0)}_{n_{m-2}}
\Phi^{(0)}_{n_{m-1}}\Phi^{(0)}_{n_m}) + ...)\bar{W}_0^{n-2} + \nonumber\\&&
+(\partial_{n_1}..\partial_{n_{m-2}}(\Phi^{(0)}_{n_{m-1}}\Phi^{(0)}_{n_m}) +
 ...)
\bar{W}_0^{n-1}
+\frac{1}{n}\partial_{n_1}...\partial_{n_{m-1}}
 \Phi^{(0)}_{n_m}\bar{W}_0^n]\stop
\label{wwi}
\eeqar
A general term in (\ref{wwi}) is, up to a combinatorial factor,
of the form
\beq
res[\bar{W}_0^{n+1-p}\prod_{i \in S}\pa_{n_i}\bar{W}_0
\prod_{j_1 \in S_1}\pa_{n_{j_1}}\bar{W}_0...
\prod_{j_n \in S_n}\pa_{n_{j_n}}\bar{W}_0] \comma
\label{gt}
\eeq
where the sets $S,S_1..S_n$ are disjoint, possibly empty, and satisfy
$S \cup S_1..\cup S_n = (1...m)$. $p$ is the number of empty sets.
The general term (\ref{gt}) is the analog in the $W_{1+\infty}$
Ward identities to (\ref{gtp}) in the degeneration equation.

Using (\ref{part}) we get
\beqar
&&\langle T_{-n}\prod_{i = 1}^m T_{n_i}\rangle  =
\frac{\Gamma(n)}{\Gamma(n - m + 2)}
- \nonumber\\&&
\frac{\Gamma(n)}{\Gamma(n - m + 3)}
\sum_{j,k = 1;j \neq k}^m (n + \sum_{j,k \neq i = 1}^m n_i)\Theta(-n -
\sum_{j,k \neq i = 1}^m n_i)
\langle T_{n + \sum_{j,k \neq i = 1}^m n_i }T_{n_j}T_{n_k}\rangle  -...
\nonumber \\&&
-
\frac{\Gamma(n)}{\Gamma(n - k + 1)}
\sum_{i_1..,i_k = 1;i_j\neq i_l}^m (n_{i_1}+..+ n_{i_k})\Theta(n_{i_1}+..+
 n_{i_k})
\langle T_{n + n_{i_1} + ..+ n_{i_k}}\prod_{i_1..i_k\neq j=1}^m
T_{n_j}\rangle \nonumber \\&&
-...-
(n-1) \sum_{i,j = 1}^m (n + n_i + n_j)\Theta(-n - n_i - n_j)
\langle T_{n + n_i + n_j}\prod_{k \neq i,j}^m T_{n_k}\rangle  -
\nonumber \\&&
\sum_{i,j = 1;i \neq j}^m (n + n_i + n_j)\Theta(-n - n_i - n_j)
(n_i + n_j)\Theta(n_i + n_j)
\langle T_{n + n_i + n_j}\prod_{i,j \neq k=1}^m T_{n_k}\rangle
\nonumber \\&&
-\sum_{i = 1}^m (n + n_i)\Theta(-n - n_i)
\langle T_{n + n_i}\prod_{i\neq j=1}^m T_{n_j}\rangle  \stop
\label{WWI}
\eeqar
The last term in (\ref{wwi}) is proportional to $\Theta(\sum_{i = 1}^m n_i)$
and it vanishes since $\sum_{i = 1}^m n_i = - n$ is negative.

The Ward identity (\ref{WWI}) is identical to the $(1,q=-1)$
degeneration equation
(\ref{tcorrel}), thus establishing the required equivalence.
This result parallels the equivalence between the Virasoro constraints
and the Verlinde's recursion relations for topological gravity\cite{VV},
and more generally between the Montano-Rivlis
degeneration equation\cite{MR} for $(1,q)$ model and $W_q$ constraints.
We have shown the equivalence at genus zero and will return to higher
genera at section 8.

\section{Discrete states as gravitational descendants}

\subsection{The spectrum of $(1,q=-1)$ model versus $2d$ string theory}

The spectrum of $2d$ string theory consists of the tachyon as well as
discrete states, i.e. states at special quantized values of
momentum\cite{Lian-KPol}. As we saw, in the topological framework the positive
momentum tachyons are most naturally identified with primary fields. The
non-positive momentum tachyons correspond to additional fields introduced into
the calculational scheme of minimal topological models and are for $q > 1$ the
auxiliary fields analogous to the primary fields. This however does not
characterize them as primaries or descendants in the topological phase of the
$2d$ string. It has been argued \cite{HOP,GM} that negative momentum tachyons
are
actually gravitational descendants of the zero momentum tachyons.
The natural conjecture is to identify the discrete states as gravitational
descendants of positive momentum tachyons \cite{HOP,GM}.

We will take this viewpoint in the following and use the degeneration
equation analytically continued to $q=-1$ in order to compute their
correlators. In analogy with the tachyons, we identify positive momentum
discrete states as gravitational descendants and negative ones as their
auxiliary analogs.
Thus,
\beq
{\cal Y}^+_{J,m} \rightarrow \frac{P_{k,\alpha}}{\alpha+kq},~~~
\alpha = J+m, k=J-m \comma
\label{dsgd}
\eeq
with $J,m = 0,\frac{1}{2},1..$ and $0 \leq m \leq J$, while
${\cal Y}^+_{J,-m}$ are considered as their auxiliary analogs.
This picture implies that the operators $\frac{P_{n=kq+\alpha}}
{\alpha +kq}$ should decouple
for $k > \alpha$ at $q=-1$, since negative momentum
discrete states are identified as the auxiliary analogs
of the positive momentum ones.
This is indeed the case as we will verify in
the next section.

\subsection{Correlators of $(1,q=-1)$ gravitational descendants}

Our aim is to establish a consistent scheme for calculating correlators
of discrete states viewed as gravitational descendants of $(1,q=-1)$
theory by analytically continuing the degeneration
equation with gravitational
descendants of $(1,q)$ minimal models to $q=-1$.
In order to pose the rules for the analytical continuation
let us perform some sample calculations.


Consider first the two-point function $\langle {\cal Y}^+_{J_1,m_1}
{\cal Y}^+_{J_2,m_2}\rangle $. Using  (\ref{cgh}) and (\ref{metric}) we
expect
that the only
non-vanishing two-point functions are
$\langle {\cal Y}^+_{J,m}{\cal Y}^+_{J,-m}\rangle $.
However, since we do not derive this result
from the degeneration equation we cannot exclude the possibility that
other two-point functions that satisfy momentum conservation do
not vanish.
The choice of the non-vanishing two-point functions
with gravitational descendants is basically
part of the definition of the $(1,q=-1)$ theory.
Different choices do not affect the correlators of the tachyons as
they do not appear in their degeneration equation.
They may show up, however, in the $(1,q=-1)$
degeneration equation for gravitational
descendants. Since we consider the latter as an analytical
continuation of the $q>1$ equation in which those correlators
do not appear, they will actually have no effect on the higher point
functions of descendants.

The simplest correlator for which we have to use the degeneration
equation is the three-point function. Thus consider the correlator
$\langle P_nP_{n_1}P_{n_2}\rangle $,
where, for instance, $P_{n_1}$ and $P_{n_2}$ are primaries while
$P_n$ is a descendant
with $n = kq + \alpha$. Such a correlator corresponds in $q=-1$ to
$\langle {\cal Y}^+_{J,m}T_{n_1}T_{n_2}\rangle $ with
$J=\frac{\alpha + k}{2}, m = \frac{\alpha - k}{2}$.
There are two different ways to compute this correlator using the
degeneration equation (\ref{3pte}) since  one can choose the marked
operator to be either the descendant or one of the primary fields.
Evidently, the result must be independent of the choice.
In the $q>1$ degeneration equation scheme an operator has as many contacts
as his primary index.
Using (\ref{3pt}) we see that the value of the correlator is
$\frac{\alpha n_1n_2}{q}$ when $P_n$ is the marked operator and
$\frac{nn_1n_2}{q}$ when one of the primary fields is  chosen as the
marked operator. The result is explicitly dependent on the choice and
consistency requires that the correlator vanishes.
Indeed the ghost number conservation law (\ref{cgh}) reads in this case
\beq
kq+\alpha + n_1 + n_2 = q + 1 \comma
\eeq
and thus $k$ must be zero.
The same problem with consistency occurs if we take two different
descendants as marked operators.
Thus there is no non-vanishing
three point function with
a descendant in $(1,q)$ models.
In order to get a consistent scheme at $q=-1$ via analytical
continuation of the $q > 1$ degeneration equation
we must require that three point
functions with discrete states vanish, although momentum conservation
does not forbid such correlators. Thus we conclude that
\beq
\langle {\cal Y}^+_{J_1,m_1}{\cal Y}^+_{J_2,m_2}{\cal Y}^+_{J_3,m_3}\rangle  =
0
 \comma
\label{ds3pt}
\eeq
unless all the three fields are tachyons.

Consider now the  four point function $\langle P_nP_{n_1}P_{n_2}P_{n_3}\rangle
 $,
where $P_n$
is a descendant and the rest are primaries.
The ghost number conservation fixes $n = q + \alpha$, that is $P_n$
is the first descendant of the primary field $P_{\alpha}$.
Using (\ref{e4pt}) we get:
\beq
\langle P_{q+\alpha}P_{n_1}P_{n_2}P_{n_3}\rangle  = \frac{\alpha(\alpha +
 q)n_1n_2n_3}
{q^2}  \stop
\label{ds4pt}
\eeq
Unlike the correlator of four primaries, the correlator (\ref{ds4pt})
does not include $\Theta$ terms and thus does not depend on
the choice of the kinematic region.

The constraint $n = q + \alpha$ on the descendant as a consequence of the
ghost number conservation law is also necessary for consistency.
If the constraint is not satisfied, that is $n = kq + \alpha, k>1$,
the value of the correlator depends on the choice of the marked operator.
In such a case, when the descendant is chosen as the marked operator we get
for the correlator
$\frac{\alpha(\alpha-1)n_1n_2n_3}{q^2}$ while if one of the primaries,
for instance $P_{n_1}$, is
taken to be the marked operator we get
$\frac{n_1(n_1-1)(kq+\alpha)n_2n_3}{q^2}$.

Consistency of the analytical continuation to $q=-1$ requires that the only
non-vanishing four point functions with discrete states are
\beq
\langle {\cal Y}^+_{J,J-1}T_{n_1}T_{n_2}T_{n_3}\rangle  = 2J-1 \comma
\label{ds4ptt}
\eeq
where we factored out the $\delta-$function enforcing the
momentum conservation $2(J-1) + n_1+n_2+n_3 =0$ and used the normalization
(\ref{dsgd}).

In our picture
${\cal Y}^+_{1,0}$ corresponds to $P_{q+1}$ which is the dilaton operator,
i.e. the first descendant of the puncture. Thus, for $J=1$ equation
(\ref{ds4ptt}) is a special case of the dilaton equation on the sphere.
Let us derive the general genus $g$ dilaton equation. Thus, consider
the genus $g$ correlator $\langle P_{q+1}\prod_{i=1}^sP_{n_i}\rangle _g$.
The degeneration equation reads:
\beq
(-q)\langle P_{q+1}\prod_{i=1}^sP_{n_i}\rangle _g +
\sum_{i=1}^s\langle \overbrace{P_{q+1}P_k}\prod_{i\neq j=1}^sP_{n_j}\rangle _g
\langle P_{-k}P_{n_i}\rangle _0 = 0
\stop
\label{dileq}
\eeq
Using (\ref{cont}) and (\ref{cgh}) we get
\beq
\langle P_{q+1}\prod_{i=1}^sP_{n_i}\rangle _g =
\frac{q+1}{q}(s+2g-2)\langle \prod_{i=1}^sP_{n_i}\rangle _g  \stop
\label{qdil}
\eeq

At $q=-1$ we have
\beq
\langle {\cal Y}^+_{1,0} \prod_{i=1}^s{\cal Y}^+_{J_i,m_i}\rangle _g =
(2-2g-s)\langle \prod_{i=1}^s{\cal Y}^+_{J_i,m_i}\rangle _g  \stop
\label{deq}
\eeq
Equation (\ref{deq}) is expected from the topological viewpoint
since the dilaton measures the
Euler characteristic of the Riemann surface with $s$ punctures, that is
$2-2g-s$.

The correlator (\ref{ds4pt}) belongs to a family of correlators that
do not include $\Theta$ terms, namely
$\langle \prod_{i=1}^n P_{n_i}\rangle , n_i = k_iq + \alpha_i$ such that
$\sum_{i=1}^n k_i = n-3$. They read
\footnote{We checked this formula analytically as well as numerically,
but we do not have a full proof for it.}
\beq
\langle \prod_{i=1}^nP_{n_i}\rangle = {1\over q^{n-2}}
{(n-3)!\over \prod_{i=1}^n k_i!} \prod_{i=1}^n\prod_{j_i=0}^{k_i}
(j_iq+\alpha_i) \stop
\label{sptype}
\eeq

Note that, as expected from the picture described in the previous section,
the correlator for the normalized operators
$\frac{P_{n_i}}{\alpha_i + k_iq}$
vanishes if $k_i > \alpha_i$.
As each descendant $P_{n = kq +\alpha}$
in a correlator seems to contribute a multiplicative factor
\beq
\prod_{j=0}^{k}(jq+\alpha) \comma
\eeq
it implies the vanishing of the correlator at $q=-1$ in theses cases.

At $q=-1$ (\ref{sptype}) corresponds to
\beq
\langle \prod_{i=1}^n{\cal Y}^+_{J_i,m_i}\rangle  =  (-1)^n
(n-3)!\prod_{i=1}^n \frac{\prod_{j_i=0}^{J_i-m_i-1}
(J_i+m_i-j_i)}
{(J_i-m_i)!} \comma
\eeq
where $\sum_{i=1}^n(J_i-m_i) = n-3$.

The first correlator with discrete states that exhibits dependence
on the choice of kinematic region is the five point.
The ghost number selection
rule allows various possibilities. We can have in the correlator
one descendant field $P_n$ with $k=1,2$ or two descendants with $k=1$.
Two of these possibilities satisfy the condition $\sum_{i=1}^5k_i = 2$ and thus
belong to the family of correlators that do not have $\Theta$ terms
as described above and are given by (\ref{sptype}).
The third possibility is the correlator $\langle P_n\prod_{i=1}^4P_{n_i}\rangle
 $ where
$n = q + \alpha$ and $P_{n_i}$ are primaries.
We can compute this correlator using the degeneration equation (\ref{5pt})
in two ways, by choosing either the descendant or one of the primaries as
the marked operator.
Taking the descendant field as the marked operator yields
\beqar
&&\langle P_n\prod_{i=1}^4P_{n_i}\rangle  =
\frac{\alpha}{q}\sum_{i=1}^4n_i\langle P_{\alpha + n_i -1}
\prod_{i\neq j=1}^4P_{n_j}\rangle  - \nonumber\\&&
\frac{\alpha(\alpha-1)}{q^2} \sum_{i,j=1; i\neq j}^4n_in_j
\Theta(\alpha + n_i + n_j - q - 2)
\langle P_{\alpha + n_i + n_j - q - 2}
\prod_{i,j\neq k=1}^4P_{n_k}\rangle  +\nonumber\\&&
\prod_{i=1}^4n_i\frac{\alpha(\alpha-1)(\alpha-2)}{q^3} \stop
\label{ds5pt}
\eeqar
It is straightforward to verify that one gets the same result if one of the
primary fields is taken as the marked operator.
Analytical continuation of (\ref{ds5pt}) to $q=-1$ proceeds in the same
manner as that of the tachyon correlators, namely
$\Theta(x)\rightarrow \Theta(-x)$. We get
\beqar
&&\langle {\cal Y}^+_{J,J-1}\prod_{i=1}^4T_{n_i}\rangle  =
-\sum_{i=1}^4\frac{(2J-1)(n_i + 2J - 2)}{2J-2}\langle T_{2(J-1)+n_i}
\prod_{i\neq j=1}^4 T_{n_j}\rangle  - \nonumber\\&&
\sum_{i,j = 1; i\neq j}^4(2J-1)(n_i + n_j +2J-2)
\Theta(- n_i - n_j -2J+2)\nonumber\\&&
\langle T_{2(J - 1) + n_i + n_j}\prod_{i,j\neq k=1}^4T_{n_k}\rangle  -
(2J-1)(2J - 3) \stop
\label{cds5pt}
\eeqar

Let us turn now to the general case. Consider the correlator
$\langle \prod_{i=1}^nP_{n_i}\rangle $ where $n_i = k_iq + \alpha_i$.
{}From the ghost number conservation law one gets the constraint
\beq
\sum_{i=1}^n k_i \leq n-3 \stop
\label{constr}
\eeq
This constraint (\ref{constr}) is also necessary for the
consistency of the whole
calculational scheme. Consider, for instance, the case where $k_1 \geq n-2$
and all the other $k_i$ zero. If we take $P_{n_1}$ as the marked operator
we get by induction:
\beq
\langle \prod_{i=1}^nP_{n_i}\rangle  =
 \frac{(-1)^{n-1}}{q^{n-2}}\prod_{i=2}^nn_i
\frac{\Gamma(\alpha_1 + 1)}{\Gamma(\alpha_1 - n + 3)} \stop
\label{npts}
\eeq
On the other hand, if we take one of the primary fields for instance $P_{n_2}$
we get
\beq
\langle \prod_{i=1}^nP_{n_i}\rangle  = \frac{(-1)^{n-1}}{q^{n-2}}\prod_{i\neq
 2}^nn_i
\frac{\Gamma(n_2 + 1)}{\Gamma(n_2 - n + 3)} \comma
\label{onpt}
\eeq
and it clearly differs from (\ref{npts}).

Thus, although the constraint (\ref{constr}) does not follow from
momentum conservation in the $q=-1$ model we must impose it in the
analytical continuation scheme in order to have a consistent framework.
An immediate consequence of the constraint is that there is no non-vanishing
$n-$ point function with more than $n-3$ descendants.
Evidently, the constraint is necessary for consistency but is it sufficient?.
For the $q > 1$ models the degeneration
equation is equivalent to the $W-$ constraints and thus to the KdV
calculational scheme and is therefore consistent.
We strongly believe that the analytic continuation process to $q=-1$
does not spoil this property. It is indeed the case for all the computations
that we made but we do not have a complete proof for that.

Finally, let us consider the Ward identity for the momentum one tachyon
$T_1$.
Topological and integrable evidence imply that it
corresponds to the puncture operator in the
topological phase of $2d$ string theory \cite{HOP}.
Consider the genus $g$ correlator $\langle P_1\prod_{i=1}^sP_{n_i}\rangle _g$
The degeneration equation at $q > 1$ reads
\beq
(-q)\langle P_1\prod_{i=1}^sP_{n_i}\rangle _g +
\sum_{i=1}^s\langle \overbrace{P_1P_k}\prod_{i\neq j=1}^sP_{n_j}\rangle _g
\langle P_{-k}P_{n_i}\rangle _0 = 0 \comma
\eeq
or
\beq
\langle P_1\prod_{i=1}^sP_{n_i}\rangle _g =
\frac{1}{q}\sum_{i=1}^sn_i\Theta(n_i+1-(q+1))
\langle P_{n_i+1-(q+1)}\prod_{i\neq j=1}^sP_{n_j}\rangle _g
\stop
\label{qpeq}
\eeq
Thus, the puncture operator shifts the momentum of a gravitational
descendant and does not affect a primary.
At $q=-1$ we have
\beqar
&&\langle T_1\prod_{i \in S_1}{\cal Y}^+_{J_i,m_i}
\prod_{j\in S_2}{\cal Y}^+_{J_j,-J_j}\prod_{k \in
S_3}{\cal Y}^+_{J_k,J_k}\rangle _g
= \nonumber\\&&
\sum_{l\in S_1}(J_j-m_l-1)\langle {\cal Y}^+_{J_l-\frac{1}{2},m_l+\frac{1}{2}}
\prod_{l\neq i\in S_1}{\cal Y}^+_{J_i,m_i}
\prod_{j\in S_2}{\cal Y}^+_{J_j,-J_j}
\prod_{k\in S_3}{\cal Y}^+_{J_k,J_k}\rangle _g
\nonumber -\\&&
\sum_{l\in S_2}(2J_l-1)\langle {\cal Y}^+_{J_l-\frac{1}{2},-J_l+\frac{1}{2}}
\prod_{i\in S_1}{\cal Y}^+_{J_i,m_i}
\prod_{l \neq j\in S_2}{\cal Y}^+_{J_j,-J_j}
\prod_{k\in S_3}{\cal Y}^+_{J_k,J_k}\rangle _g \comma
\label{T1WI}
\eeqar
with $S_1\cup S_2\cup S_3 = (1...s)$.

Considering (\ref{T1WI}) as the puncture equation in the topological
phase of $2d$ string theory implies that indeed the negative momentum
tachyons and all the discrete states correspond to gravitational
descendants while non-negative momentum tachyons correspond to gravitational
primaries.
Note, however, that we cannot take a negative tachyon to be the marked
operator
within the degeneration equation framework while considering him as
a gravitational descendant of the zero momentum tachyon, since it will
have zero number of contacts. On the other hand we saw that we can take it
to be the marked operator with a formal negative number of contacts.
The latter is probably related to the fact that gravitational
descendants can be described via matter degrees of freedom
\cite{Lossev}.

\section{ $(1,q=-1)$ topological recursion
relations via one and two splittings}

In \cite{LaSo} it was shown that correlation functions of any $(1,q>1)$
model
can be computed using
recursion relations derived  not from the $W_q$ constraints
but rather from the   $W_3$ constraints on the partition function.
In terms of the topological
procedure this means that only terms with one and two splittings of the
Riemann surface are needed in order to reproduce the
results of the $(1,q)$ degeneration equation.

It is  natural to ask whether a similar statement
can be made about the topological procedure for the $q=-1$ model.
As we shall see in this section, this is
indeed the case. This will provide us with new  recursion relations for
tachyon correlation functions, as well as a support for the
 consistency of the $(1,q=-1)$ degeneration equation for gravitational
descendants.

Let us first briefly summarize the method of \cite{LaSo} for the $q>1$
minimal topological models. The idea is to write down an  algorithm for
  computing genus $g$ correlators
 $\langle  \prod_{i=1}^{N}
P_{k_i,\alpha_i} \rangle _g $ based only on one and two splittings.
Consider the following  relations
\begin{eqnarray}
&&\langle  P_{0,2}^{\alpha_1-2} P_{k_1+\alpha_1-2,2} \prod_{i=2}^N
P_{k_i,\alpha_i} \rangle _g
 = c_1 \langle  \prod_{i=1}^N P_{k_i,\alpha_i}
\rangle _g + \Delta_1
\nonumber\\&&
\langle  P_{k_1+\alpha_1-2,2} P_{0,2}^{\alpha_1-2} \prod_{i=2}^N
P_{k_i,\alpha_i} \rangle _g
 = c_2 \langle  \prod_{i=1}^N P_{k_i,\alpha_i}
\rangle _g + \Delta_2   \comma
\label{mishw3}
\end{eqnarray}
where  $ \alpha_1 = min\{ \alpha_i \} $ and  $c_1$ and $c_2$ are some
coefficients.   $\Delta_1$ and $\Delta_2$ are each a sum of correlators
which follow from the contacts of $P_{0,2}$ and  $P_{k_1+\alpha_1-2,2}$
respectively.
The particular choice of the left hand sides (l.h.s) of equations
(\ref{mishw3})
is made so that the following  properties are obeyed:
(i) The degeneration equation
of  only one and two splittings
is required; (ii) $c_1-c_2\ne 0$;   (iii)
all the terms in $\Delta_1$ and $\Delta_2$ are either
on genus $g'< g$ or on genus $g$ and
contain less operators then $N$ or contain $N$ operators with
$ min\{ \alpha_i \} < \alpha_1$.
The second property was proven by invoking certain novel polynomial  identities
which are generalizations of Abel's identity \cite{LaSo} while the third one
was shown to hold using induction procedure.
By
subtracting the two equations (\ref{mishw3}) we derive the
following recursion
relation
\begin{equation}
\langle  \prod_{i=1}^N P_{k_i,\alpha_i} \rangle _g  =
{\Delta_2-\Delta_1\over c_1-c_2}\label{mishcc}  \stop
\end{equation}
We would like to emphasize that since  $\Delta_1$ and $\Delta_2$
include only ``lower correlators", (\ref{mishcc}) is a recursion algorithm
for determining the original correlator.
Does this procedure hold also for the $q=-1$ description
of the $c=1$ string model?. It seems that since in the  derivation of
(\ref{mishcc}) we have used only the degeneration  equation together with the
consistency of the calculational  scheme,  it should hold also for the
$(1,q=-1)$
model.
 Let us examine things more carefully.
Unlike the $q>1$ cases,  at $q=-1$,$c_1$ which is given by \cite{LaSo}
\beq
c_1=\Bigl ( {2 \over q}
\Bigr )^{\alpha_1-2} \prod_{i=1}^{\alpha_1-2} [(k_1+i)q+2] \comma
\eeq
vanishes  for certain values of $k_1$ and $\alpha_1$:
 $k_1=0$, $\alpha_1 \ge 4$ and $k_1=1$, $\alpha_1 \ge 3$ .
Note, however,
that $c_2$ is different from zero. The
difference $c_1-c_2$ is independent of $q$ and is given
by \cite{LaSo}
\beq
c_1-c_2=2^{\alpha_1-2} \prod_{i=1}^{\alpha_1-2} (k_1+i) \stop
\eeq
This obviously implies that even for the cases where $c_1=0$
the recursion relation  (\ref{mishcc}) still holds:
$\langle  \prod_{i=1}^N P_{k_i,\alpha_i} \rangle _g  =
-{\Delta_2\over c_2}$.

We have thus found that for the whole range of values of $k_1,\alpha_1$ one can
express the correlators  in terms of ``lower correlators" using
the degeneration equation  with only one and two splittings,
thus obtaining a new set of ``Ward identities".

The main advantage of these new ``Ward identities" for $q>1$ is that
 they provide a practical algorithm to
compute correlators on higher genus Riemann surfaces.
Let us briefly explain why the topological procedure of \cite{MR} fails in
these cases.
When inserting (\ref{id}) in a degeneration of the
sphere, no more than one field of (\ref{id}) contributes to the summation
due to the ghost number conservation.
On higher genus surfaces things change.
When pinching a handle we insert both fields of (\ref{id})
on the same surface, thus the ghost number is conserved for
an infinite number of fields.
The $\Theta$ functions
that are inserted in order to avoid contributions from
unphysical operators restrict the sum and prevent infinities.
Two point functions, however,  do not include these
$\Theta$ functions and therefore divergences due
to the infinite sum occur. For $q=2$ the
regularization for these infinities is known, and
for $q=3$ no new infinities occur.
Thus, the new set of ``Ward identities"
provide a regularized method for calculating correlators on higher
genus Riemann surfaces.
For the case of interest $q=-1$ this algorithm enables us to compute
correlators of gravitational descendants
on $g>0$ surfaces. Furthermore one may
use it to deduce the ``postulated regularization" which we use in the next
section to retrieve the genus one tachyon
correlators derived from the $W_{1+\infty}$ Ward identities.
In the appendix an explicit evaluation of $<P_4P_6P_{-10}>$ on the sphere
as well as $<P_4P_{-4}>$ on the torus are presented

\section{$(1,q=-1)$ model on higher genus Riemann surfaces}

We have seen that at genus zero, that the topological recursion
relations for tachyon correlators implied by the degeneration equation
analytically continued to the $(1,q=-1)$ theory coincide with the
$W_{1+\infty}$ matrix model Ward identities of the $c=1$ string.
In this section we will show that the equivalence between these
two sets of Ward identities persists beyond genus zero.
Since the topological recursion relations for genus greater than
zero receive contributions from pinching of handles, infinities
are encountered due to the propagation of infinite number of fields in the
degeneration
and should be properly regularized.
We will prove explicitly the equivalence for the genus one case,
and discuss the equivalence for genus $g$ surfaces.

\subsection{Tachyon correlators on the torus via $(1,q=-1)$ theory}

Let us begin by computing the two and three point tachyon correlators
on the torus using the degeneration equation and compare to the
$W_{1+\infty}$ Ward identities.

Consider first the correlator for two tachyons on the torus
$\langle T_nT_{-n}\rangle _1$.
The degeneration equation reads:
\beqar
&&\langle P_nP_{-n}\rangle _1 +
n\langle \overbrace{P_nP_i}\rangle _1\langle P_{-i}P_{-n}\rangle _0 +
\sum_{i \geq 0} (n-1)\langle \overbrace{P_nP_iP_{-i}}P_{-n}\rangle _0
 \nonumber\\&&
+ \sum_{i \geq 0} (n-1)(n-2)\langle \overbrace{P_nP_jP_iP_{-i}}\rangle _0
\langle P_{-j}P_{-n}\rangle _0  = 0  \stop
\label{g12pt}
\eeqar
The first two terms come from splitting and the rest from pinching of
the handle as shown in Fig.4. Note that in the latter terms
the marked operator contacts two operators inserted around the
pinching point. This implies that one has to insert $n-2$ additional splitted
spheres.

\vskip 0.5cm
\putfig{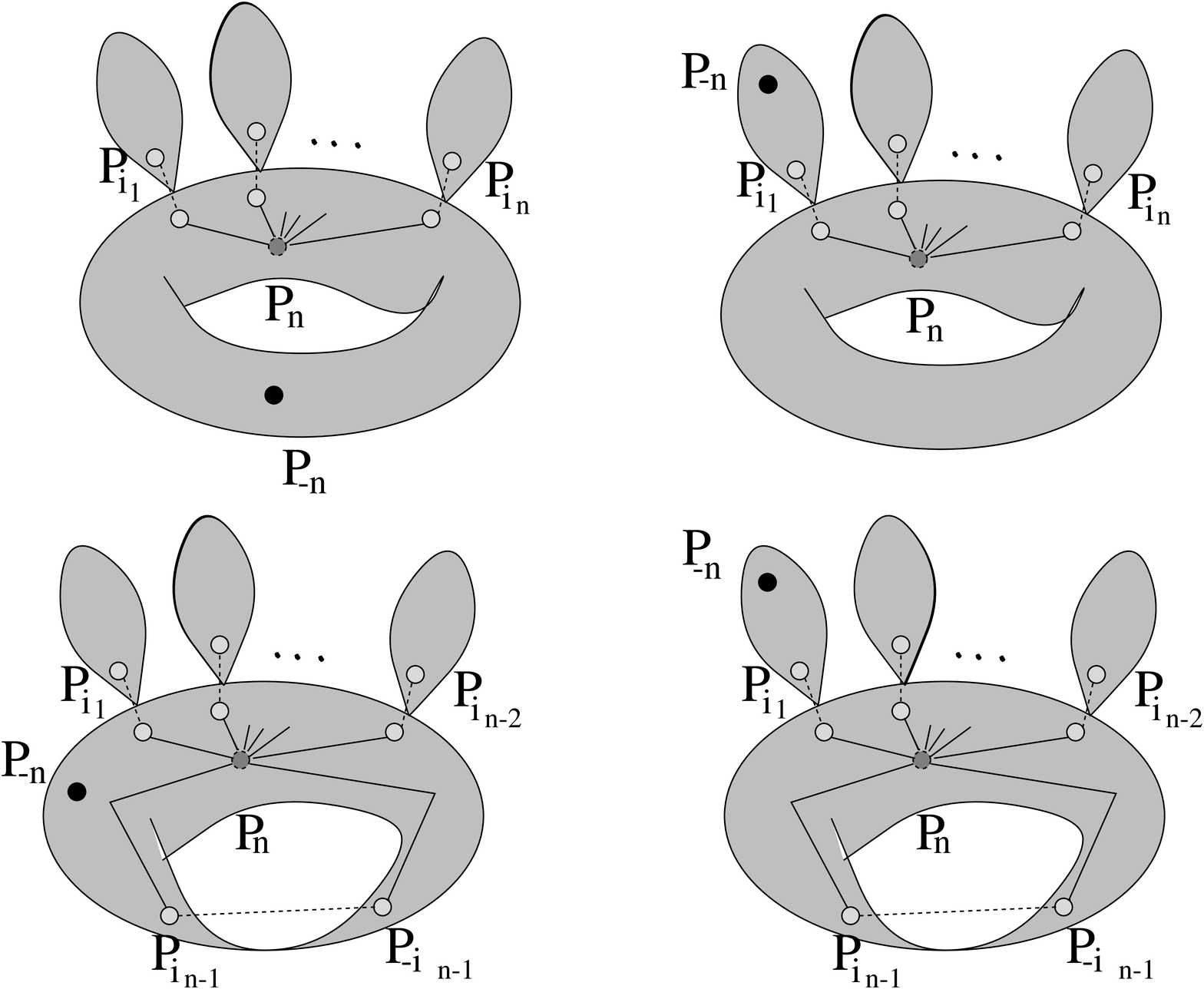}{ Fig. 4: The degeneration equation for two-point
function on the torus}{100mm}

The second term vanishes since it is proportional
to $\langle  P_{-q-1} \rangle _1 \equiv 0$, while
the last two terms need to be regularized.
We will now postulate a regularization for the contact
operation which we justify, as we will show,
by that it yields the correct results for general
tachyon correlators on the torus. We also checked for several cases
that it reproduces the
correct results for minimal models.

The regularization reads:
\beqar
&&reg[\sum_{i \geq 0}
\langle \overbrace{P_nP_iP_{-i}\prod_{j=1}^mP_{n_j} }\rangle _0]
\equiv \frac{n}{24}
\comma \nonumber\\&&
reg[\sum_{i \geq 0}
\langle \overbrace{P_nP_iP_{-i}\prod_{j=1}^mP_{n_j} } P_{n_{m+1}}\rangle _0]
\equiv \frac{n}{24}(1+ n_{m+1})(2+ n_{m+1})\langle  P_{-n_{m+1}}
P_{n_{m+1}} \rangle _0  \comma
\nonumber\\&&
reg[\sum_{i \geq 0}
\langle \overbrace{P_nP_iP_{-i}\prod_{j=1}^mP_{n_j} } \prod_{j=m+1}^s
 P_{n_j}\rangle _0]
\equiv \frac{n}{24}(1+\sum_{j=m+1}^s n_j)(2+ \sum_{j=m+1}^s n_j)
\nonumber\\&&
\Theta(\sum_{j=m+1}^s n_j)
\langle P_{n+ \sum_{j=1}^m n_j}\prod_{j=m+1}^sP_{n_j}\rangle _0 \stop
\label{reg}
\eeqar

Using the regularized contact
(\ref{reg}) in equation (\ref{g12pt}) we get
\beq
\langle T_nT_{-n}\rangle _1 = \frac{1}{24}(n-1)(n+1)(n-2) \comma
\label{nnone}
\eeq
which is the correct result at the self-dual radius \cite{KL}.
As in the genus zero case we have a sign difference between the tachyon
correlators derived from the $q=-1$ degeneration equation and those
that were derived from the matrix model.
The second example in the appendix demonstrates how using the one and two
splitting method one  finds an identical result to that  derived from the
postulated regularization.

In order to see the  reasoning behind the definition of the
regularized contact consider the  $W_{1+\infty}$ Ward identities
for the two-point function on the torus.
Using (\ref{WIw}) we get
\beqar
&&\langle T_nT_{-n}\rangle _1 = res[\bar{W}_0^{n-1}\pa_{-n}\bar{W}_0
\bar{W}_1 + \frac{1}{n}\bar{W}_0^n\pa_{-n}\bar{W}_1
\nonumber\\&&
-\frac{1}{24}(n-1)
\bar{W}_0^{n-2}\pa_{-n}\bar{W}_0^{''}
-\frac{1}{24} (n-1)(n-2)\bar{W}_0^{n-3}\pa_{-n}\bar{W}_0\bar{W}_0^{''}]
\stop
\label{Wt2pt}
\eeqar
The second term in (\ref{Wt2pt}) vanishes, as can be seen from (\ref{Wexp}).
The other terms of (\ref{Wt2pt}) coincide with those of (\ref{g12pt})
if we use (\ref{Wpart}) and
the definition of the regularized contact (\ref{reg}).

Consider next the three point function of tachyons on the torus
$\langle T_nT_{n_1}T_{n_2}\rangle _1$, where $n+n_1+n_2=0$.
The degeneration equation reads :
\beqar
&&\langle P_nP_{n_1}P_{n_2}\rangle _1 + (n\langle
 \overbrace{P_nP_i}P_{n_2}\rangle _1\langle P_{-i}P_{n_1}\rangle _0
+ (n_1\leftrightarrow n_2)) \nonumber\\&&
+ ((n-1)(n-2)\langle \overbrace{P_nP_iP_{-i}P_j}P_{n_2}\rangle _0\langle
 P_{-j}P_{n_1}\rangle _0
+ (n_1\leftrightarrow n_2)) + \nonumber\\&&
(n-1)(n-2)(n-3)\langle \overbrace{P_nP_iP_{-i}P_jP_k}\rangle _0\langle
 P_{-j}P_{n_1}\rangle _0
\langle P_{-k}P_{n_2}\rangle _0 = 0 \comma
\label{tr3pt}
\eeqar
where we have written only the non-vanishing terms.
The first three terms in (\ref{tr3pt}) arise from splitting and the
rest from pinching.
We can use now the regularization (\ref{reg}) together with the appropriate
$\Theta$ terms and get, up to a sign,
the right three point function for tachyons on the
torus \cite{KL}
\beq
\langle T_nT_{n_1}T_{n_2}\rangle _1 =
 \frac{1}{24}(n-1)(n-2)(n_1^2+n_2^2-n-2)\stop
\eeq
It is more instructive, however, to compare the degeneration
equation (\ref{tr3pt}) with the $W_{1+\infty}$ Ward identities for the
three point function.
Using (\ref{WIw}) we get:
\beqar
&&\langle T_nT_{n_1}T_{n_2}\rangle _1 = res[\bar{W}_0^{n-1}(\pa_{n_1}\bar{W}_0
\pa_{n_2}\bar{W}_1 +(n_1\leftrightarrow n_2)) - \frac{1}{24} (n-1)(n-2)
\bar{W}_0^{n-3} \nonumber\\&&
(\pa_{n_1}\bar{W}_0\pa_{n_2}\bar{W}_0^{''} + (n_1\leftrightarrow n_2))
-\frac{1}{24}(n-1)(n-2)(n-3)
\bar{W}_0^{n-4}\bar{W}_0^{''}\pa_{n_1}\bar{W}_0\pa_{n_2}\bar{W}_0]
\comma
\label{Wt3pt}
\eeqar
where, similarly to equation (\ref{tr3pt}), we have written only the
non-vanishing terms.
Comparing (\ref{tr3pt}) and (\ref{Wt3pt}) we see that by associating
the pinching of a handle with $W''$, namely
\beqar
&&\sum_{i \geq 0}\langle \overbrace{P_nP_iP_{-i}P_jP_k}\rangle _0
 \leftrightarrow
\frac{n}{24}\bar{W}_0^{''} \nonumber\\&&
\sum_{i \geq 0}\langle \overbrace{P_nP_iP_{-i}P_j}P_k\rangle _0 \leftrightarrow
\frac{n}{24}\pa_k \bar{W}'' \comma
\label{wpp}
\eeqar
and using  (\ref{Wexp}), (\ref{Wpart}) the two equations are identical.
The generalization of (\ref{wpp}) is clear,
\beq
\sum_{i \geq 0}\langle \overbrace{P_nP_iP_{-i}\prod_{j\in S}P_{n_j}}
\prod_{k\in S'}P_{n_k} \rangle_0 \leftrightarrow \frac{n}{24}
\prod_{k\in S'}\pa_{n_k}\bar{W}_0^{''} \stop
\eeq

\subsection{The equivalence between the $(1,q=-1)$  degeneration
equation and the $W_{1+\infty}$ Ward identities on the torus}

Our aim is to prove for the torus that the topological recursion relations of
the $(1,q=-1)$ theory are
identical to the $W_{1+\infty}$ Ward identities of $2d$ string theory.

Consider the correlator $\langle T_{n}\prod_{i = 1}^m T_{n_i}\rangle _1$.
Using the Ward identities we have
\beqar
&&\langle T_{n}\prod_{i = 1}^m T_{n_i}\rangle  \equiv
 \partial_{n_1}...\partial_{n_m}
\langle \langle T_{n}\rangle \rangle (t=0) =
\frac{1}{n} \partial_{n_1}...\partial_{n_m} res(\bar{W}_0^{n}\bar{W}_1) -
\nonumber\\&&
\frac{1}{24}(n-1) \partial_{n_1}...\partial_{n_m} res(\bar{W}^{n-2}_0
\bar{W}''_0) \stop
\label{npt}
\eeqar
We will distinguish between two types of terms, those that include a genus
one correlator and arise topologically from splittings and those that
include only genus zero correlators and arise topologically from
pinching the handle of the torus.
The former are derived from the first term in (\ref{npt}) and the latter
from the second term.

Consider first the terms coming from the process of splitting.
They read:
\beqar
&&res[\sum_{i=1}^m [\frac{\Gamma(n)}{\Gamma(n - m + 2)}\Phi_{n_1}^{(0)}..
\Phi_{n_i}^{(1)}..\Phi_{n_m}^{(0)}\bar{W}^{n - m + 1}_0 + \nonumber\\&&
\frac{\Gamma(n)}{\Gamma(n - m + 3)}(\partial_{n_1}(\Phi_{n_2}^{(0)}
..\Phi_{n_i}^{(1)}..\Phi_{n_m}^{(0)})
+ \Phi_{n_1}^{(0)}\partial_{n_2}(\Phi_{n_3}^{(0)}
..\Phi_{n_i}^{(1)}..\Phi_{n_m}^{(0)}) + \nonumber\\&&
+.. + \Phi_{n_1}^{(0)}..
\Phi_{n_i}^{(1)}..
\partial_{n_{m-1}} \Phi_{n_m}^{(0)})\bar{W}^{n-m+2}_0 + \nonumber\\&&
+...+
\frac{\Gamma(n)}{\Gamma(n - k + 1)}(\partial_{n_1}...\pa_{n_{m-k-1}}
(\Phi_{n_{m-k}}^{(0)}..\Phi_{n_i}^{(1)}..
\Phi_{n_m}^{(0)}) + ...)\bar{W}^{n-k}_0 + ...] \nonumber\\&&
+ (n-1)(\partial_{n_1}..\partial_{n_{m-3}}(\Phi_{n_{m-2}}^{(1)}
\Phi_{n_{m-1}}^{(0)}\Phi_{n_m}^{(0)}) + ...)\bar{W}^{n-2}_0 + \nonumber\\&&
+(\partial_{n_1}..\partial_{n_{m-2}}
(\Phi_{n_{m-1}}^{(1)}\Phi_{n_m}^{(0)}) + ...)
\bar{W}^{n-1}_0
+\frac{1}{n}\partial_{n_1}...\partial_{n_{m-1}} \Phi_{n_m}^{(1)}
\bar{W}^n_0]\stop
\label{twwi}
\eeqar
This has the same structure as (\ref{wwi}) with one of the $\Phi$ being
$\Phi^{(1)}$ and the rest being $\Phi^{(0)}$.
A general term in (\ref{twwi}) takes the form of (\ref{gt}) with
one of the $\bar{W}_0$ replaced by $\bar{W}_1$.
Using (\ref{part}) we get
\beqar
&&\langle T_{-n}\prod_{i = 1}^m T_{n_i}\rangle _1 = \sum_{i=1}^m [
\frac{\Gamma(n)}{\Gamma(n - m + 2)}n_i\langle T_{n_i}T_{-n_i}\rangle _1
- \nonumber\\&&
\frac{\Gamma(n)}{\Gamma(n - m + 3)}(
\sum_{i \neq j = 1}^m (n + \sum_{i,j \neq l = 1}^m n_l)\Theta(-n -
\sum_{i,j \neq l = 1}^m n_l)
\langle T_{n + \sum_{i,j \neq l = 1}^m n_l }T_{n_i}T_{n_j}\rangle _1
- \nonumber \\&&
\sum_{i \neq j,k = 1}^m (n + \sum_{j,k \neq l = 1}^m n_l)\Theta(-n -
\sum_{j,k \neq l = 1}^m n_l)n_i
\langle T_{n + \sum_{j,k \neq l = 1}^m n_l }T_{n_j}T_{n_k}\rangle _0
\langle T_{n_i}T_{-n_i}\rangle _1) - \nonumber\\&&
... -
\frac{\Gamma(n)}{\Gamma(n - k + 1)}
\sum_{i \neq i_1..,i_k = 1}^m (n_{i_1}+..+ n_{i_k})
\Theta(n_{i_1}+..+ n_{i_k})n_i \nonumber\\&&
\langle T_{n + n_{i_1} + ..+ n_{i_k}}\prod_{i_1..i_k \neq j=1}^m T_{n_j}\rangle
 _0
\langle T_{n_i}T_{-n_i}\rangle _1 - \nonumber \\&&
-...-
(n-1) \sum_{i \neq j,k = 1}^m (n + n_j + n_k)\Theta(-n - n_j - n_k)n_i
\langle T_{n + n_j + n_k}\prod_{j,k \neq l=1}^m T_{n_l}\rangle _0
\nonumber \\&&
\langle T_{n_i}T_{-n_i}\rangle _1 ...-(n + n_i)\Theta(-n - n_i)n_i
\langle T_{n + n_i}\prod_{i\neq j =1}^m T_{n_j}\rangle _0\langle
 T_{n_i}T_{-n_i}\rangle _1]\stop
\label{tWWI}
\eeqar

Consider now the correlator $\langle P_n\prod_{i = 1}^m P_{n_i}\rangle _1$,
 where $P_n$
is a primary operator. Taking $P_n$ as the marked operator, the
degeneration equation reads:
\begin{eqnarray}
&&(-q)^n\langle P_n\prod_{i = 1}^m P_{n_i}\rangle _1 + n(-q)^{n-1}
\sum_{i = 1}^m \langle P_{n + n_i-(q+1)}
\prod_{i,j = 1;i\neq j}^m P_{n_j}\rangle _1
\langle P_{-n_i}P_{n_i}\rangle _0+ \nonumber\\&&
n(-q)^{n-2} \sum_{i \neq j = 1}^m
\langle P_{n + n_i + n_j-2(q+1)}\prod_{i,j \neq k = 1}^m P_{n_k}\rangle _1
\langle P_{-n_i -n_j +(q+1)}P_{n_i}P_{n_j}\rangle _0 +
\nonumber\\&&
n(n-1)(-q)^{n-2} \sum_{i,j = 1;i\neq j}^m
\langle P_{n + n_i + n_j -2(q+1)}\prod_{i,j \neq k = 1}^m P_{n_k}\rangle _1
\langle P_{-n_i}P_{n_i}\rangle _0 \langle P_{-n_j}P_{n_j}\rangle _0 +...
\nonumber\\&&
+ \frac{\Gamma(n + 1)}{\Gamma(n - k + 1)}(-q)^{n-k}
\sum_{i_1..,i_k = 1}^m
\langle P_{n + n_{i_1} + ..+ n_{i_k}-k(q+1)}
\prod_{i_1,..,i_k \neq i = 1}^m P_{n_i}\rangle _1
\prod_{l = 1}^k \langle P_{-n_{i_l}}P_{n_{i_l}}\rangle _0
\nonumber\\&&
+ ... + \frac{\Gamma(n + 1)}{\Gamma(n - m + 3)}(-q)^{n-m+2}
\sum_{j,k=1;j\neq k}^m
\langle P_{n + \sum_{j,k \neq i=1}^m n_i - (m-2)(q+1)}P_{n_j}P_{n_k}
\rangle _1\nonumber\\&&
\prod_{j,k\neq i=1}^m\langle P_{-n_i}P_{n_i}\rangle _0
+.. + \frac{\Gamma(n + 1)}{\Gamma(n - m + 2)}(-q)^{n-m}
\sum_{i=1}^m\prod_{i \neq j = 1}^m \langle P_{-n_i}P_{n_i}\rangle _1
\langle P_{-n_j}P_{n_j}\rangle _0 \stop
\label{tqcorrel}
\end{eqnarray}
A general term in (\ref{tqcorrel}) takes the form of (\ref{gtp}) with one
of the correlators is on the torus and the rest on the sphere.
Passing to $q=-1$ it is easy to see that to each term in
(\ref{tWWI}) correspond an identical term in (\ref{tqcorrel}) and
vice versa.

Let us turn now to the contributions from surfaces with pinched handles.
The degeneration equation yields
\beqar
&&(n-1)(n-2)\sum_{i \geq 0}\langle \overbrace{P_nP_{n_j}P_iP_{-i}}\prod_{j \neq
 k
 =1}^mP_{n_k}\rangle _0
\langle P_{n_j}P_{-n_j}\rangle _0 + \nonumber\\&&
(n-1)(n-2)(n-3)\sum_{i \geq 0}\langle \overbrace{P_nP_{n_j}P_{n_k}P_iP_{-i}}
\prod_{j,k \neq l =1}^mP_{n_l}\rangle _0
\langle P_{n_j}P_{-n_j}\rangle _0 \langle P_{n_k}P_{-n_k}\rangle _0
\nonumber\\&&
+...
\comma
\label{pinch}
\eeqar
with a general term depicted in Fig.5

\vskip 0.5cm
\putfig{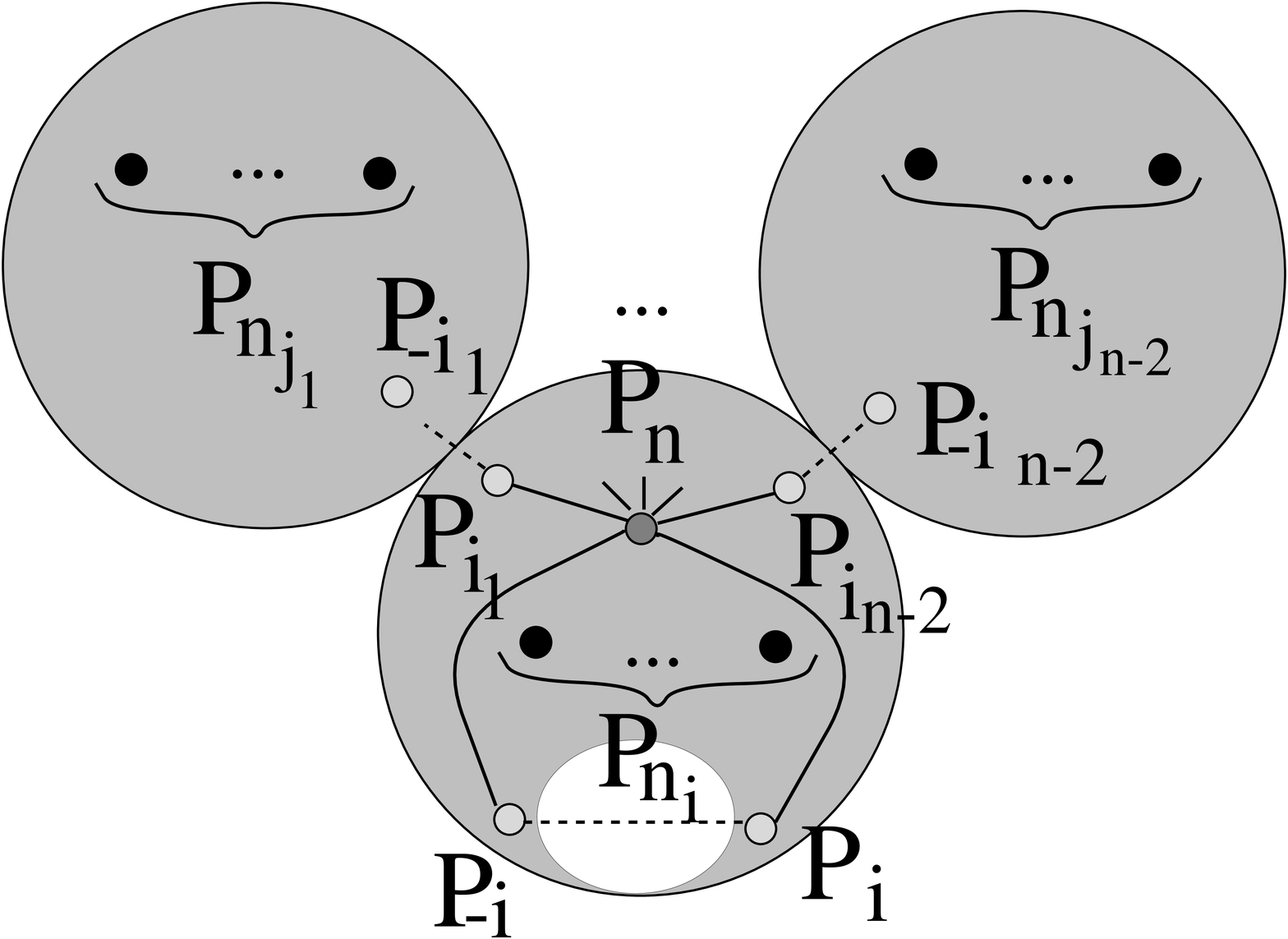}{ Fig. 5: Pinching a handle in the genus one
degeneration equation}{100mm}

\beq
\sum_{i\geq 0}\langle \overbrace{P_nP_{i_1}...P_{i_{n-2}}P_iP_{-i}}
\prod_{i \in S}P_{n_i}\rangle _0
\langle P_{-i_1}\prod_{j_1 \in S_1}P_{n_{j_1}}\rangle _0...
\langle P_{-i_{n-2}}\prod_{j_{n-2} \in S_{n-2}}P_{n_{j_{n-2}}}\rangle _0 \comma
\label{gttp}
\eeq
where the sets $S,S_1..S_{n-2}$ are disjoint, possibly empty, and satisfy
$S \cup S_1..\cup S_{n-2} = (1...m)$.

The corresponding terms in the $W_{1+\infty}$ Ward identities are
\beqar
&&\frac{1}{24}res[(n-1)(n-2)\pa_{n_j}\bar{W}_0\prod_{j \neq k =1}^m\pa_{n_k}
\bar{W}_0^{''} +\nonumber\\&&
(n-1)(n-2)(n-3)\pa_{n_j}\bar{W}_0\pa_{n_k}
\bar{W}_0\prod_{j,k \neq l =1}^m\pa_{n_l}
\bar{W}_0^{''} + ...]
\comma
\label{wpinch}
\eeqar
with a general term of the form
\beq
\frac{1}{24}res[\prod_{i \in S}\pa_{n_i}\bar{W}_0^{''}
\prod_{j_1 \in S_1}\pa_{n_{j_1}}\bar{W}_0...
\prod_{j_{n-2} \in S_{n-2}}\pa_{n_{j_{n-2}}}\bar{W}_0] \stop
\label{gwt}
\eeq

Using  (\ref{reg}) and (\ref{wpp}) we establish the equivalence of equations
(\ref{pinch}) and (\ref{wpinch}).

\subsection{The $(1,q=-1)$ degeneration equation versus $W_{1+\infty}$
Ward identities for genus $g$ Riemann surfaces}

Consider a general genus $g$ correlator $\langle P_n\prod_{i=1}^mP_{n_i}\rangle
 _g$
in $(1,q=-1)$ theory.
The degeneration equation for this correlator receives contributions
, as depicted in Fig.6,
from splittings, for instance
\beq
\langle P_{n+ \sum_{i \in S}n_i}\prod_{j \in S'}P_{n_j}\rangle _g
\langle P_{-n_{i_k}-n_{i_l}}P_{n_{i_k}}P_{n_{i_l}}\rangle _0..  \comma
\label{split}
\eeq
with $S \cup S'= (1...m)$,
from pinching a dividing cycle such as
\beq
\langle P_{n + \sum_{i \in S}n_i}\prod_{j \in S'}P_{n_j}\rangle _{g-g_1}
\langle P_{-n_{i_k}-n_{i_l}}P_{n_{i_k}}P_{n_{i_l}}\rangle _{g_2}..  \comma
\label{dcycle}
\eeq
and from pinching $h$ non-trivial homology cycles
\beq
\sum_{j_1...j_h \geq 0}
\langle \overbrace{P_{n+\sum_{i \in S}n_i}
P_{j_1}P_{-j_1}...P_{j_h}P_{-j_h}}
\prod_{j \in S'}P_{n_j}\rangle _{g-h}
\langle P_{-n_{i_k}-n_{i_l}}P_{n_{i_k}}P_{n_{i_l}}\rangle _0..  \stop
\label{hcycle}
\eeq
\vskip 0.5cm
\putfig{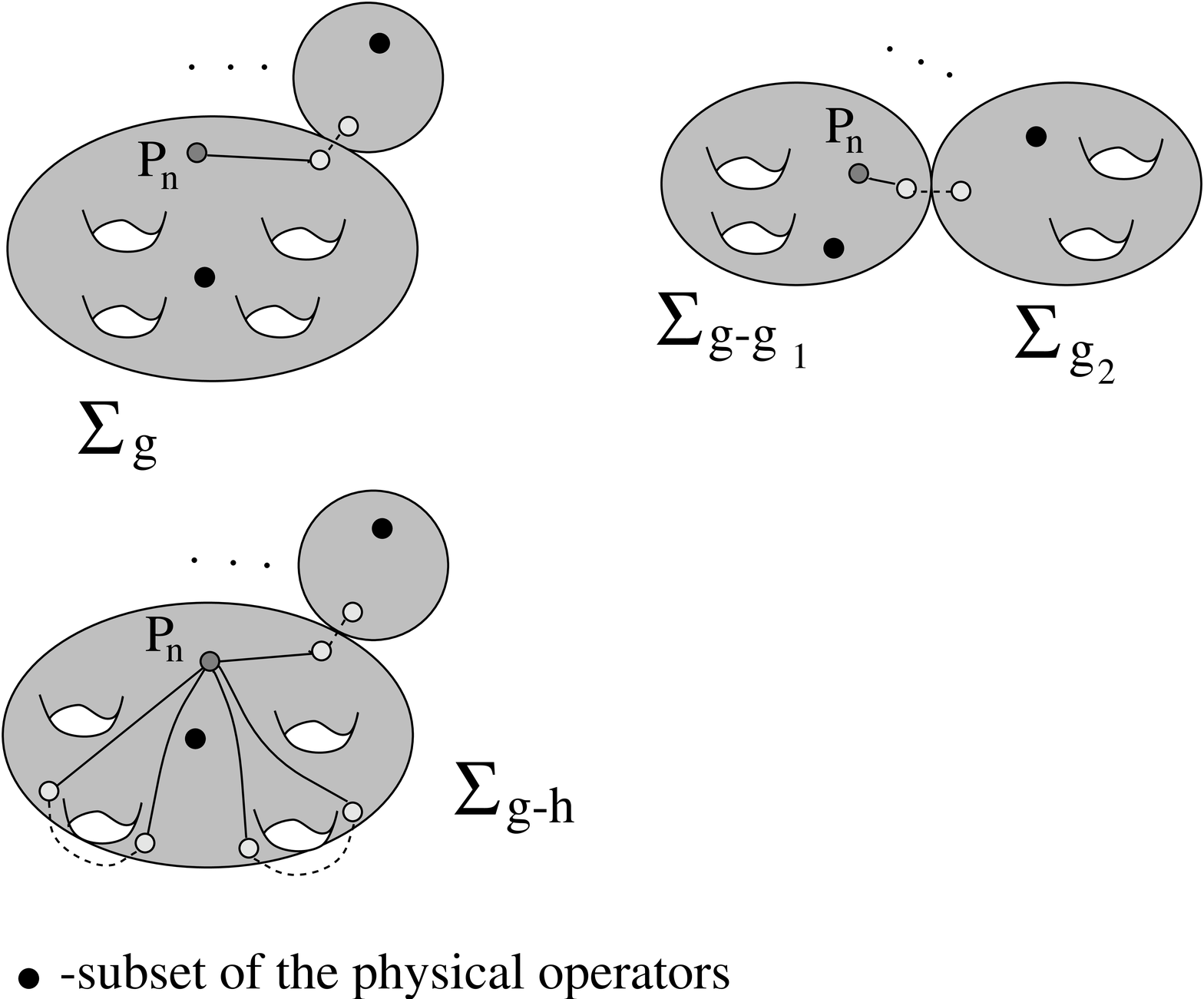}{ Fig. 6: Contributions to the genus $g$ degeneration
equation}{100mm}

Consider now a genus $g$ tachyons correlator $\langle
 T_n\prod_{i=1}^mT_{n_i}\rangle _g$.
{}From the $W_{1+\infty}$ Ward identities we expect terms with a similar
structure to (\ref{split}), (\ref{dcycle}) and (\ref{hcycle}).
The term that corresponds to the splitting (\ref{split}) is
\beq
res[\prod_{i \in S'}\pa_{n_i}\bar{W}_g
\pa_{n_{i_k}}\pa_{n_{i_l}}\bar{W}_0... ]
\comma
\label{wsplit}
\eeq
the term that corresponds to the pinching of a dividing cycle  (\ref{dcycle})
is
\beq
res[\prod_{i \in S'}\pa_{n_i}\bar{W}_{g-g_1}
\pa_{n_{i_k}}\pa_{n_{i_l}}\bar{W}_{g_2}...]
\comma
\label{wcycle}
\eeq
and the term that corresponds to the pinching of a non-trivial cycle
(\ref{hcycle})
\beq
res[\prod_{i \in S'}\pa_{n_i}\bar{W}_{g-h}^{(2h)}
\pa_{n_{i_k}}\pa_{n_{i_l}}\bar{W}_0...]
\stop
\label{whcycle}
\eeq
Evidently, the contribution from reduction of the genus by pinching
a non-trivial homology cycle needs to be properly
regularized in the framework
of the degeneration equation in order to be identified with (\ref{whcycle}).

The general structures of the degeneration equation and the
$W_{1+\infty}$ Ward identities for any genus seem to be equivalent,
thus it is plausible to conjecture that they coincide.
However more work is needed in order to prove this in details.

\section{ Summary and Conclusions}

We studied in this paper the  $(1,q=-1)$ model coupled to topological
gravity. This model is supposed to describe $2d$ string theory at the
self-dual radius, that is at its topological phase.
We defined the model by analytical continuation of the
$q>1$ degeneration equation to $q=-1$.
We have shown that at genus zero the $q=-1$ degeneration equation yields
the genus zero $W_{1+\infty}$ Ward identities for tachyon correlators.
The positive momentum tachyons were identified as primary fields while
the negative ones as their analytically continued $q>1$
auxiliary analogs.
By defining the discrete states of $2d$ string theory as the gravitational
descendants of $(1,q=-1)$ model we proposed a scheme for computing
their correlators and computed some of them.
Unfortunately we could not compare the results
since these correlators have not been computed by other
means.
We derived the puncture equation as well as the dilaton equation,
with the puncture operator being the momentum one tachyon, and the dilaton
operator being, as usual, its first descendant.
The puncture equation supports the
identification of the negative momentum tachyons
as well as the discrete states as gravitational descendants.

We showed that in a similar manner to the $q>1$ models,
there exist recursion relations for the correlators that consist of
only one and two splittings. Unlike the $W_{1+\infty}$ Ward identities
which consist only of tachyon fields, the new recursion relations
involve $(1,q=-1)$ gravitational descendants.
This provides us with other means to compute
tachyon correlators and supports the consistency of the whole
computational scheme.
However, an open problem is to fully prove the consistency of the
$(1,q=-1)$ degeneration equation for gravitational descendants.
A related question is what is the algebra underlying the $(1,q=-1)$
degeneration equation. For $q>1$ the algebra is $W_q$, while for $q=-1$
on the space of tachyon times the algebra is $W_{1+\infty}$,
but the algebra on the full phase space that includes also $q=-1$
gravitational descendants is not known.
It is plausible to conjecture that it is also a $W_{1+\infty}$
algebra.

As we discussed in section 7,
on $g>0$ Riemann surfaces one encounters infinities in the degeneration
equation framework that should be regularized. A proper
regularization is known for the $q=2,3$ models, that is for
one and two splittings of the Riemann surface.
Thus, when using the recursion relations (\ref{mishcc})
analytically continued to $q=-1$, we can calculate any genus $g$
correlator of primaries and descendants.
However, the nature of this procedure is numerical and it does not easily
yield closed and explicit expressions for correlators.
One example of this type is described in the appendix.
We therefore took another route. We
postulated a regularized contact and used it to prove that
the $q=-1$ degeneration equation for tachyon correlators coincides
with the $W_{1+\infty}$ Ward identities on the torus.
Derivation of the regularized contact from the recursion relations
(\ref{mishcc}) is important and may lead to a
regularized degeneration equation for $q=-1$ gravitational descendants.
This is currently in study.
As we showed, the structure of the  $q=-1$ degeneration
equation for tachyon correlators seems to parallel
that of the $W_{1+\infty}$ Ward identities for any genus.
It is important to show that this is indeed correct in details, since
it will provide us with a topological interpretation of the $W_{1+\infty}$
Ward identities and thus with another link between the topological
and integrable structures of $2d$ string theory.

There are several open problems that require further study.
First, there are specific questions related to the present work.
(i) Is the analytical continuation of the $q>1$ models to $q=-1$
unique? (ii) Can one prove that
the $q=-1$ model, as described in this paper, indeed
describes $2d$ string theory
at the self-dual radius without the need to compare all
correlation functions calculated
in the present approach to those derived via other methods.

Second, there are questions related to more general implications\\
1. The relation between the degeneration equation and intersection
theory  for $q>1$ as well as for $q=-1$.\\
2. Can we consistently analytically continue the degeneration equation
to other domains of definition of $q$ such as complex values, and  are
 there corresponding physical systems.\\
3. Is it possible to write topological recursion relations based on
contacts and degenerations for any topological string theory?.
Computations done in \cite{BCOV} seem to fall into this category.\\
4. Finally, other generalizations defined by the assignment of different
charges, e.g. with non-abelian structure, and different anomalous
conservation laws may lead to topological recursion
relations describing interesting theories.

\appendix{Recursion relations via one and two splittings, a numerical
example}

In this appendix we present two  numerical computations
of  correlators using the topological procedure of section seven, namely,
 via iterative use of
one and two splittings of the Riemann surface.

Consider first the correlator $\langle P_4 P_6 P_{-10}\rangle _0$  which
corresponds
 to
the tachyons three point function on the sphere.
The degeneration equation for this correlator consists of at least
four splittings, which is the case when we choose $P_4$ as the
marked operator. Its numerical value at $q=-1$, before normalization,
is 240 as given by (\ref{3pt}).
Following (\ref{mishw3}), let us consider the correlator
$\langle P_{2,2} P_2P_2P_6P_{-10}\rangle _0$, which vanishes by (\ref{sptype}).
Choosing $P_{2,2}$ to be the marked operator we have
a degeneration equation with one and two splittings
\beqar
&&0 \equiv \langle P_{2,2} P_2P_2P_6P_{-10}\rangle _0 = {2 \over q}[4\langle
P_2
 P_{1,3} P_6
P_{-10}\rangle _0 +6\langle P_{1,7} P_2 P_2 P_{-10}\rangle _0 \nonumber\\&&
- 10\langle P_{1,-9} P_2 P_2 P_6\rangle _0]
-{2 \over q^2}[4\langle P_4 P_6 P_{-10}\rangle _0+
12\langle P_2 P_8P_{-10}\rangle _0
\nonumber\\&&
- 40\langle P_2 P_6 P_{-8}\rangle _0
-60\langle P_2 P_2 P_{-4}\rangle _0]
\nonumber\\&&
+{2 \over q}[\langle P_2 P_2 P_{-4}\rangle _0\langle P_4 P_6 P_{-10}\rangle _0
+ 2\langle P_2 P_6P_{-8}\rangle _0\langle P_2 P_8 P_{-10}\rangle _0]    \comma
\label{ex1}
\eeqar
where we used (\ref{cont}) and (\ref{2pt}).
Note that we do not have $\Theta$ terms in (\ref{ex1}).
The reason is that the operator $P_{2,2}$ can perform a contact with two
physical operators and the result is still physical:
\beq
\overbrace{P_{2,2}P_nP_m} = P_{n+m}  \stop
\eeq
Since the analytical continuation is defined such that we allow
in the correlators operators $P_{n}$ with $n$ negative
only at the final $q=-1$ recursion relations, there are no $\Theta$
terms in a five-point function with $P_{2,2}$ as the marked operator.
Note in contrast that a five-point function with $P_{1,1}$ as
the marked operator includes $\Theta$ terms (\ref{ds5pt}).

The r.h.s of equation (\ref{ex1}) consist of the required correlator plus
other terms, all of which can be computed via one and two splittings.
Note that the required correlator for primaries is expressed via
correlators of primaries and descendants.
Using the degeneration equation
for these terms we get $\langle  P_4 P_6 P_{-10} \rangle _0 =240$ as expected.

Consider next the correlator $\langle P_4P_{-4} \rangle_1$ which
corresponds to the tachyons two-point function on the torus.
The degeneration equation for this correlator consists of four splittings,
and its value is $-20$.
Following (\ref{mishw3}), let us consider the correlator
$\langle P_{2,2} P_2P_2P_{-4}\rangle_1$ which vanishes by the arguments
of section six.
Taking $P_{2,2}$ as the marked operator we get
\beqar
&&0 \equiv \langle P_{2,2} P_2P_2P_{-4}>_1 = {2 \over q} [ 4 < P_2 P_{1,3}
P_{-4}\rangle_1
 -4\langle P_{1,-3} P_2 P_2 \rangle_1 ] \nonumber\\&&
-{2 \over q^2} [ 4\langle P_4  P_{-4}\rangle_1]
+ {2 \over q}  \langle P_2 P_2 P_{-4}\rangle_0
\langle P_4  P_{-4}\rangle_1 \nonumber\\&&
-{1 \over q^2} [-4 \langle P_2 P_2 P_{-2} P_{-2}\rangle_0
- 8 \langle P_2 P_2 P_{-1} P_{-3}\rangle_0]  \stop
\label{2pex}
\eeqar
Using $P_2$ as the marked operator for the computation of the first two
correlators on the r.h.s of (\ref{2pex}) we have
\beqar
&&\langle P_4 P_{-4} \rangle_1 = 4 \langle P_2 P_{-1} P_{-1}\rangle_0
-8  \langle P_{1,3}
P_{-2}\rangle_1 \nonumber\\&&
+{1 \over 2}  \langle P_2 P_2 P_{-2} P_{-2}\rangle_0
 +\langle P_2 P_2 P_{-1} P_{-3}\rangle_0 \stop
\label{2pex1}
\eeqar
In order to evaluate (\ref{2pex1})
we have to compute the correlator  $\langle P_{1,3} P_{-2}
\rangle_1$, which we calculate by considering the correlator
$\langle P_2P_{2,2}P_{-2} \rangle_1$. We get
$\langle P_{1,3} P_{-2}\rangle_1 =-1$, and using (\ref{2pex1})
we arrive at the required result
\beq
\langle P_4P_{-4} \rangle_1 = -20 \stop
\eeq
\vspace{2.0cm}

{\bf Acknowledgment} We would like to thank A. Hanany, A. Lossev,
R. Plesser and S. Yankielowicz
for helpful discussions.

\end{document}